\documentclass[12pt,preprint]{aastex}

\newcommand\be{\begin{equation}}
\newcommand\en{\end{equation}}

\newcommand\kms{\rm{\, km \, s^{-1}}}

\shorttitle{High-Resolution Spectroscopy in Tr37}
\shortauthors{Sicilia-Aguilar et al.}

\begin{document}

\title{High-Resolution Spectroscopy in Tr37: Gas Accretion Evolution in Evolved Dusty Disks\footnote{Observations reported here 
were obtained at the MMT Observatory, a joint
facility of the Smithsonian Institution and the University of Arizona.}}

\author{Aurora Sicilia-Aguilar\altaffilmark{2}, Lee W. Hartmann\altaffilmark{3},}
\author{G\'{a}bor F\"{u}r\'{e}sz\altaffilmark{4,}\altaffilmark{5},Thomas Henning\altaffilmark{2}}
\author{Cornelis Dullemond\altaffilmark{2}, Wolfgang Brandner\altaffilmark{2}}

\altaffiltext{2}{Max-Planck-Institut f\"{u}r Astronomie, K\"{o}nigstuhl 17, 69117 Heidelberg, Germany}
\altaffiltext{3}{University of Michigan, 830 Dennison 500 Church St., Ann Arbor, MI 48109}
\altaffiltext{4}{Center for Astrophysics, 60 Garden Street, Cambridge, MA 02138}
\altaffiltext{5}{University of Szeged, Department of Experimental Physics, Dom ter9, H-6723
Szeged, Hungary}

\email{sicilia@mpia.de}

\begin{abstract}

Using the Hectochelle multifiber spectrograph, 
we have obtained high-resolution (R$\sim$34,000) spectra in the H$\alpha$ region
for a large number of stars in the 4 Myr-old cluster Tr 37, containing 146
previously known members and 26 newly identified ones.
We present the H$\alpha$ line
profiles of all members, 
compare them to our IR observations of dusty disks (2MASS/JHK + IRAC + MIPS 24$\mu$m),
use the radial velocities as a membership criterion, and calculate
the rotational velocities. 
We find a good correlation between the accretion-broadened profiles and the
presence of protoplanetary disks, noting that a small
fraction of the accreting stars presents broad profiles with H$\alpha$ equivalent
widths smaller than the canonical limit separating CTTS and WTTS. 
The number of strong accretors appears to be lower than in younger regions, 
and a large number of CTTS have very small accretion
rates (\.{M}$\leq$10$^{-9}$M$_\odot$/yr). Taking into account that the 
spectral energy distributions are consistent with
dust evolution (grain growth/settling) in the innermost disk, 
this suggests a parallel
evolution of the dusty and gaseous components.
We also observe that about half of the ``transition objects'' 
(stars with no IR
excesses at $\lambda\leq$6 $\mu$m) do not show
any signs of active accretion, whereas the other half is accreting with 
accretion rates $\leq$10$^{-9}$M$_\odot$/yr. 
These zero or very low accretion rates reveal important gas
evolution and/or gas depletion in the innermost disk, which could be
related to grain growth up to planetesimal or even planet sizes.
Finally, we examine the rotational velocities of accreting and non accreting
stars,
finding no significant differences that could indicate disk locking at these
ages.
\end{abstract}


\section{Introduction \label{intro}}

Multiple observations in the past few decades have confirmed that 
low-mass stars are born surrounded by relatively dense and optically thick 
circumstellar disks, due to the conservation of angular momentum during the
contraction of the cores in a molecular cloud
(Lynden-Bell \& Pringle 1974; Appenzeller \& Mundt 1989; among others). 
These optically thick, accretion
disks are one of the main characteristics of very young,
pre-main sequence stars. More than 80\% of the low-mass stars aged $\sim$ 1 Myr
have accreting disks, with accretion rates around 10$^{-8}$ M$_{\odot}$/yr
(Hillenbrand et al. 1995; Haisch et al. 2001; Gullbring et al. 1998).
By the age of $\sim$4 Myr, disk fractions drop to around $\sim$45\%,
and accreting disks are a rare feature in regions aged
$\sim$10-12 Myr \citep{strom89,skrutskie90,lada00,carpenter90,armitage03,sic05ir},
suggesting that most of the disk dissipation (and probably planet formation)
occurs within 1-10 Myr \citep{haisch01,bergin04,forrest04,dalessio05,sic05ir}.
Moreover, the changes in the spectral energy distributions (SEDs) at these ages
suggest that important dust evolution (via grain growth and/or dust settling) 
is taking place during these ages (Sicilia-Aguilar et al. 2006).

According to the presence of disks and accretion,
young pre-main sequence stars are typically classified as classical
T-Tauri stars (CTTS) or weak-lined T-Tauri stars (WTTS). This differentiation 
is based on the presence or absence of accretion, normally inferred from 
strong emission lines in CTTS (mostly H$\alpha$, H$\beta$, see Herbig et al. 1996; Muzerolle et al. 1998a, 
2001), in contrast to the weak chromospheric lines 
in WTTS \citep{appenzeller89,white03}. UV observations reveal that CTTS tend to have
UV excesses that can be explained by an accretion shock \citep{calvet98,gullbring98}
and near-IR excesses that suggest the presence of dusty optically thick disks 
(Haisch et al. 2001 among others).
Therefore, CTTS (or Class II objects) are supposed to have
optically thick accretion disks, whereas WTTS (or Class III objects) are assumed to be diskless
or ``naked'' \citep{bertout89,hartigan90} or to have optically thin ``debris'' disks 
\citep{lada87,andre94}.
Nevertheless, some observations suggest that some weak-line stars may not be
``true'' WTTS since they are sometimes found to have optically thick disks 
\citep{gregoriohetem02,littlefair04}.
Recent studies taking advantage of new-generation
instruments (like the Spitzer Space Telescope) have revealed 
more of these objects in which inner dusty disks are 
dissipated and/or suffer strong dust coagulation/sedimentation
(reducing drastically the emission in near- and mid-IR), 
and accretion phenomena are not present or 
are reduced to very low rates.
These objects would belong to an intermediate and 
rapid transitional phase during which gaps open in the
disk, and accretion drops by several orders of magnitude
or ceases completely (see the cases of TW Hya, Calvet et al. 2000, Uchida et al. 2004;
and CoKu Tau/4, Forrest et al. 2004, D'Alessio et al. 2005, among
others). 
Since these ``transition objects'' are very rare, the suggested timescales
for complete dissipation/agglomeration into large grains of the rest of the
outer disk are very fast (from $\sim$ 10$^5$ yr to less than 1 Myr), 
but the timescales for gap opening, and the way accretion terminates
remain uncertain because of the lack of large samples of ``transition
objects'' at different ages and evolutionary stages.
Older regions with many transitional disks are therefore 
very interesting places to search for hints of planet formation, as this 
rapid phase could be intimately related to the formation of
planets \citep{lin86,bryden00,quillen04}.

One of the best places to find a larger sample of evolved disks and ``transition objects''
is the cluster Tr 37 in the Cep OB2 association \citep{platais98}.
Located at 900 pc distance \citep{contreras02} and aged $\sim$4 Myr
(Sicilia-Aguilar et al. 2005), Tr 37 has been studied extensively at
multiple wavelengths using some of the new-generation instruments
like the Spitzer Space Telescope and the multifiber spectrograph Hectospec
(Sicilia-Aguilar et al. 2004, 2005, 2006; from now on
Paper I, Paper II, and Paper III, respectively). Low-resolution
spectra revealed an important low-mass
population ($\sim$ 160 stars with spectral types mid-G to M2),
of which approximately 45\% show excesses characteristic of
disks (Paper III). In most cases, the presence of a 
disk is well-correlated with
the presence of accretion derived from U band photometric excesses and/or the
strong H$\alpha$ equivalent width (EW) from low-resolution spectra 
(Paper II, Paper III). Nevertheless, the disks in Tr 37 
appear more evolved than the disks observed in younger regions (i.e. Taurus),
since more than 90\% of them show a significant decrease of emission 
at shorter IR wavelengths, consistent with grain growth and/or dust settling in the innermost
disk (Paper III). This lower near-IR excess is accompanied in many cases
by a small accretion rate (presumably $\leq$10$^{-9}$M$_\odot$/yr) and/or a small H$\alpha$ EW.
In about 10\% of the total number of disks, there is no near-IR excess: 
These are the ``transition objects'', 
which do not present IR excesses at wavelength shorter than 5.8 $\mu$m
and have moreover no evidence of accretion or very low accretion rates. 
These objects are likely to have $\sim$1 AU gaps in their inner dusty
disks, but have otherwise CTTS-looking optically thick disks at larger 
distances (as inferred from mid-IR excesses at longer wavelengths; Paper III).
Therefore, Tr 37 is a unique region to study accretion in very evolved and ``transitional objects'', the end of the accretion activity, and the processes leading to gap opening.

In this context, we targeted a large number of stars in Tr 37 with the high-resolution
multifiber spectrograph Hectospec at the MMT.
Since very low accretion rates 
($\sim 10^{-9}-10^{-10}$ M$_{\odot}$/yr) are not always detectable via U band excesses (Paper II)
and produce small H$\alpha$ EW, high-resolution H$\alpha$ 
spectroscopy is probably the best method
to determine the presence of accretion in evolved disks.
High-resolution H$\alpha$ spectra reveal not only the EW of the
line, but its broadening due to magnetospheric accretion \citep{hartmann98,muzerolle98a,
muzerolle98b,muzerolle01},
providing a direct proof of accretion or non-accretion even at 
rates between 10$^{-10}$ and 10$^{-12}$ M$_{\odot}$/yr.  In addition, high-resolution
spectra can be used to obtain radial and rotational (V$sini$) velocities,
which help to refine the membership and to study
the evolution of angular momentum in older
stars, and whether the rotation 
rate is correlated to accretion, as suggested by the disk
locking theories (Shu et al. 1994; Choi \& Herbst 1996; Hartmann 2002, among others).
In Section \ref{observations} we describe the data taking and 
processing, Section \ref{results} presents the results concerning
accretion and rotation, Section \ref{discussion} contrasts these
result with out previous optical and IR observations, and we finally 
list our concluding remarks.

\section{Observations and data reduction \label{observations}}

Hectochelle is the high-resolution, multifiber echelle spectrograph,
which operates at the wide-field mode of the Cassegrain 6.5m MMT telescope in 
Mount Hopkins, AZ \citep{sze98,fabricant04}. The fiber array is
shared between the low-resolution spectrograph Hectospec, and the
high-resolution Hectochelle. Hectochelle can obtain up to
240 simultaneous spectra within a 1 degree field of view, using 
all the available 240 fibers. The fibers are 250 $\mu$m in diameter,
subtending 1.5 arcsec on the sky. 
We used the spectral order centered in H$\alpha$, which has an 
approximate width of 180 \AA\, and a resolution of R$\sim$34,000.
This is the same configuration we used in our H$\alpha$ study of
the Orion Nebula Cluster, ONC, \citep{sic05onc}.

We observed two fields in Tr 37 during two engineering runs on
December 1st, 2004 and December 2nd, 2004. A total of 231 and
233 objects were observed in the configurations, respectively.
The objects were selected giving special priority to previously
known members with apparent contradictions between the
signs of accretion (H$\alpha$ EW, U band excess; Paper I,II) 
and IR excesses consistent with disks (Paper III), followed by probable members
with noisy low-resolution spectra, and the rest of
the bona-fide members (see Paper II). A total of 146 previously known members and lower-probability members were observed.
The remaining fibers were allocated in order to observe the maximum
number of potential cluster members, which had been identified
via optical and infrared photometry, but were never observed in
our spectroscopy campaigns to determine membership (see the
procedures described in Paper II
to obtain the best candidates for spectroscopy). 
The total exposure 
time was 90 min (divided in 3 x 30 min exposures) for the 
field observed in the first night,
and 60 min (divided in 3 x 20 min exposures) for the field
observed during the second night. Due to time limitations during
the engineering run, the offset sky exposures planned (Paper II)
were too
short and noisy to be usable. Nevertheless, the subtraction of the background
nebular H$\alpha$ is possible when the stellar component is broad, as the narrow H$\alpha$ nebular emission
can be identified, fitted and subtracted \citep{muzerolle98b,sic05onc}. 
For the narrow-lined
stars, no background subtraction was possible, therefore,  
we do not use the H$\alpha$ as a requirement for
membership, establishing it via Li absorption
in our low-resolution spectra (Paper I, II), or using 
radial velocities (in the case of the new members).

The data were reduced according to standard procedures, using
IRAF\footnote{IRAF is distributed by the National Optical Astronomy
Observatories, which are operated by the Association of Universities for Research
in Astronomy, Inc., under cooperative agreement with the National 
Science Foundation} and the tasks available under the
packages \textit{mscred} and \textit{specred}. The spectra
were flattened and extracted using a dome flat and the
tasks \textit{apdefault} and \textit{apflatten}, defining
fibers and background interactively. Apertures were organized
separating odd and even numbers (due to a shift in wavelength), 
and each group was calibrated
in wavelength using ThAr comparison spectra.
As mentioned previously, no background subtraction could be
done due to the lack of appropriate offset background spectra during
this engineering run. The difficulties of subtracting the nebular
emission in highly variable H II regions prevents us from using the
sky spectra obtained with the fibers not assigned to objects
(Paper I, Paper II).
Therefore, the background subtraction for the H$\alpha$ line was done
manually in the cases where it was possible to distinguish the
background, nebular, narrow component, from the broad H$\alpha$
component (which are most of the broad-lined stars). 
The individual spectra were displayed using IRAF task \textit{splot},
and the narrow component was visually identified,
fitted to a gaussian, and subtracted. For most of the stars with
broad profiles, this procedure does not introduce significant
errors in the measurement of the EW, measured with the 
IRAF task \textit{splot}. 
In the case where the nebular H$\alpha$ was successfully subtracted,
we used \textit{splot} to measure the H$\alpha$ velocity width at
10\% of the maximum.
The H$\alpha$ EW and 10\% width, as well as the labels of
broad and narrow H$\alpha$ profiles, are given in Table \ref{spec-dat}.
Examining the 16 background spectra obtained with the non-assigned fibers, we
find that the H$\alpha$ nebular emission is always narrow and
variable in intensity, even though in a few cases
a blueshifted or redshifted narrow absorption is present, in addition to 
the emission. It is worth
to note that these absorptions could be responsible for the absorption
observed in a few of the cluster members, being especially visible in the
case of narrow-lined stars.

Finally, radial and rotational velocities for the observed stars were obtained via 
cross-correlation with similar spectral type standards observed with Hectochelle. As for
our ONC study \citep{sic05onc}, we used the IRAF task \textit{xcsao}, 
available within the \textit{rvsao} package (see Kurtz et al. 1992; 
and the \textit{xcsao} documentation in tdc-www.cfa.harvard.edu/iraf/rvsao/xcsao). 
This routine obtains the
radial velocity from the shift of the cross-correlation peak, and 
its broadening is used to calculate the rotational velocity. 
In all cases, the H$\alpha$ region and the NII lines were
removed from the spectra before the cross correlation, as we did in our
previous work on the ONC (Sicilia-Aguilar et al. 2005). The errors
depend on the signal-to-noise of the cross correlation and on the width
of the peak, and the parameter R is an indication of the goodness
of the cross-correlation and its errors \citep{tonry79,hartmann86}.
The error in the rotational velocity is proportional to a constant multiplied
by 1/(1+R) \citep{kurtz92}, and in general it
can be estimated as the rotational velocity itself divided by (1+R) \citep{hartmann86},
so values of R near 1 indicate very uncertain rotational velocities (for a more
detailed explanation of the errors for Hectochelle spectra, see F\"{u}r\'{e}sz et al.
2006).
Due to small tilts in the fibers during the construction of the spectrograph, an extra source
of error affects the wavelength calibration and, therefore, the radial velocity
measurements. The fiber tilts produce a variation in the radial velocity that
can be up to 0.6-0.8 km/s considering all the fibers, and a variation up to 
3 km/s in the V$sini$, both depending on the region of the spectrum \citep{furesz06}. 
Even though in most cases here the signal-to-noise ratio (S/N) dominates the error, 
these extra variations in the tilt may be taken into account to
explain the dispersion of the radial velocities in the cluster.
Because of the time limitations imposed during this engineering run, we could obtain
good cross-correlations and velocity estimates for about 75\% of the observed
stars (133 members and potential members) with good S/N. The radial and
rotational velocities as well as the R signal-to-noise parameter of the cluster
members and potential members are listed
in Table \ref{spec-dat}.

\section{Results\label{results}}

\begin{figure}
\epsscale{1.1}\plottwo{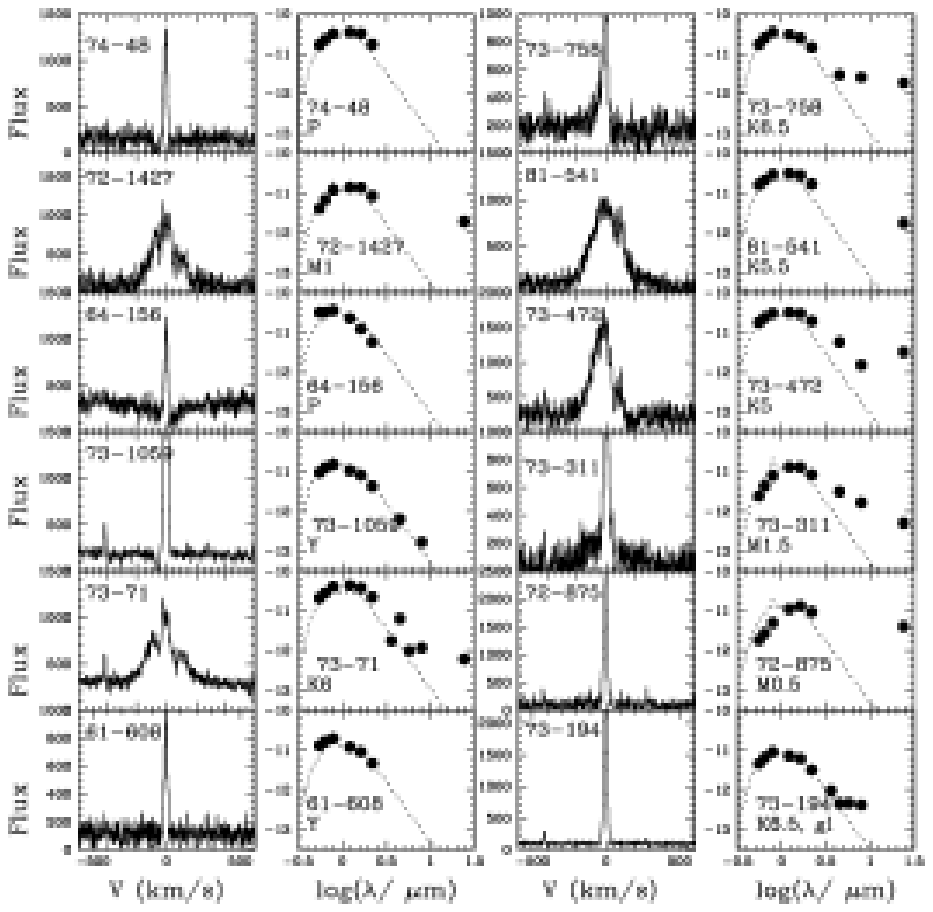}{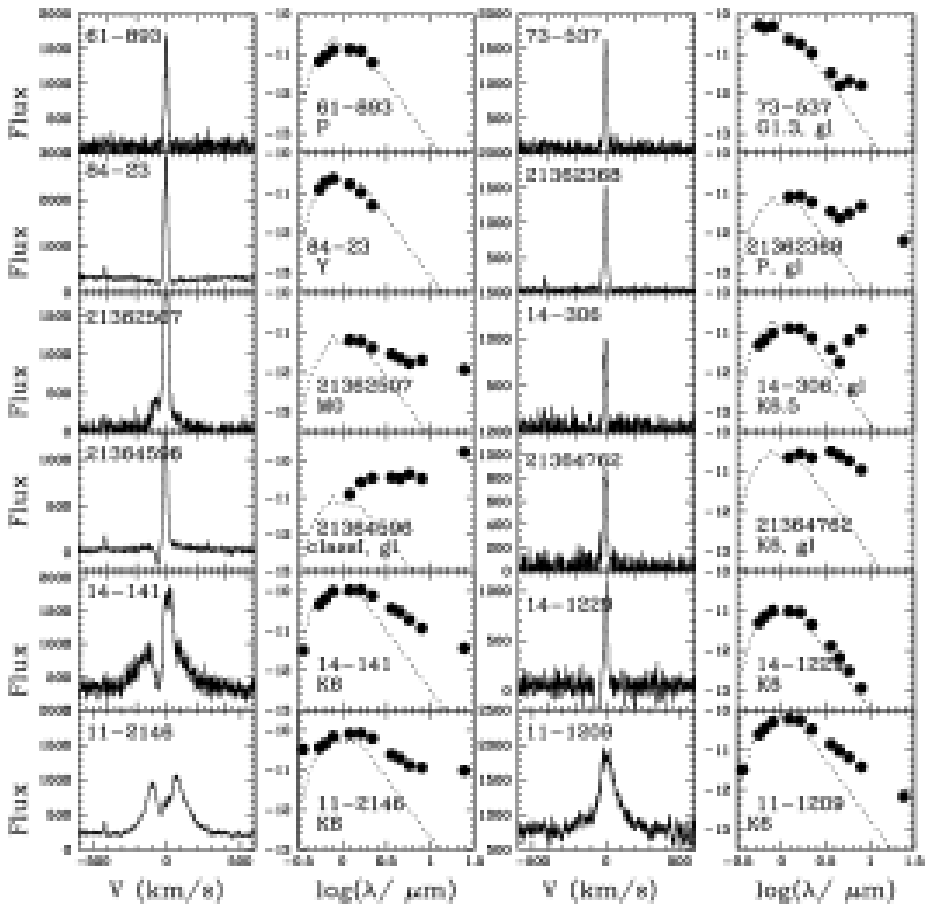}
\caption{H$\alpha$ profiles and SEDs of the observed members and potential members,
showing the optical UVRI and JHK 2MASS fluxes (Paper I, II), 
IRAC (3.6, 4.5, 5.8, 8.0 $\mu$m; Paper III) and MIPS(24$\mu$m; Paper III) data.
A photosphere \citep{kenyon95} for the given spectral type is displayed in each case
to clarify the presence of an IR excess. The 
nebular emission was successfully subtracted from the broad 
H$\alpha$ profiles with good S/N. Narrow H$\alpha$ lines, as well as some low S/N
broad profiles, are contaminated by the nebular emission.
The units of the fluxes are counts (non-calibrated) for the H$\alpha$ lines, and log($\lambda$F$_\lambda$) in
erg cm$^{-2}$ s$^{-1}$ for the SEDs. The label ``gl'' marks the stars embedded in the
Tr 37 globule, which may have large errors in the Spitzer photometry (mostly in the
5.8 and 8.0 $\mu$m IRAC bands).
 \label{profi1}}
\epsscale{1}
\end{figure}

\begin{figure}
\epsscale{1.1}\plottwo{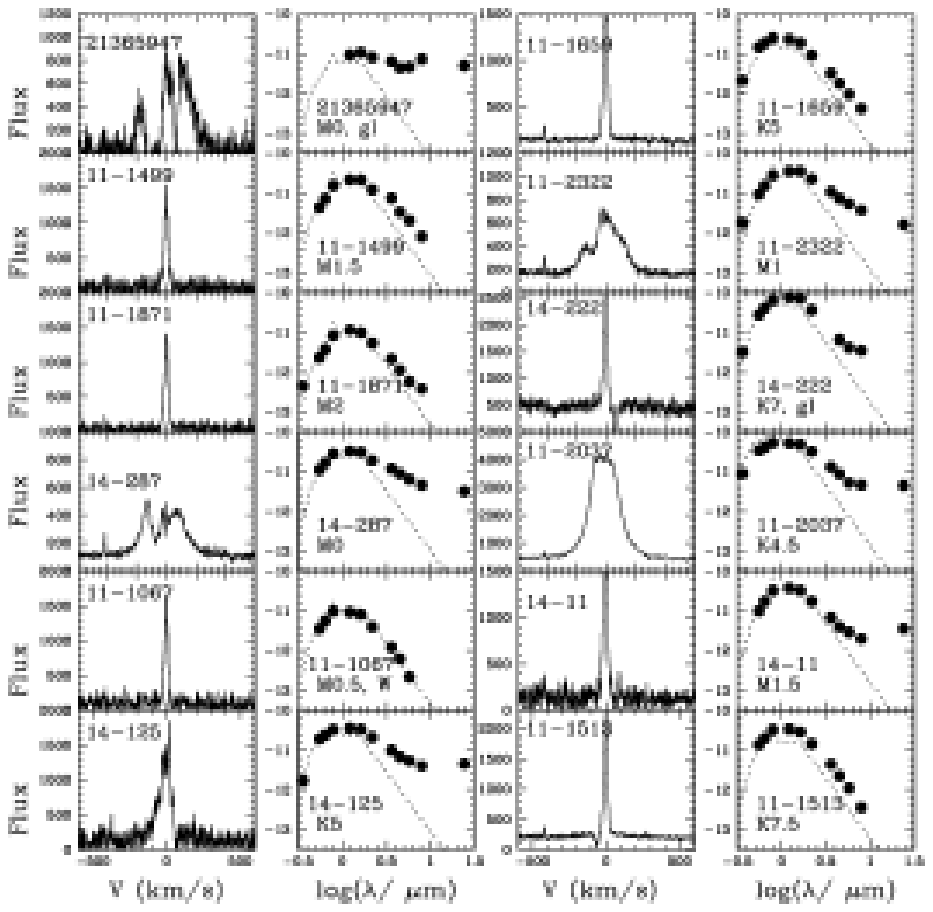}{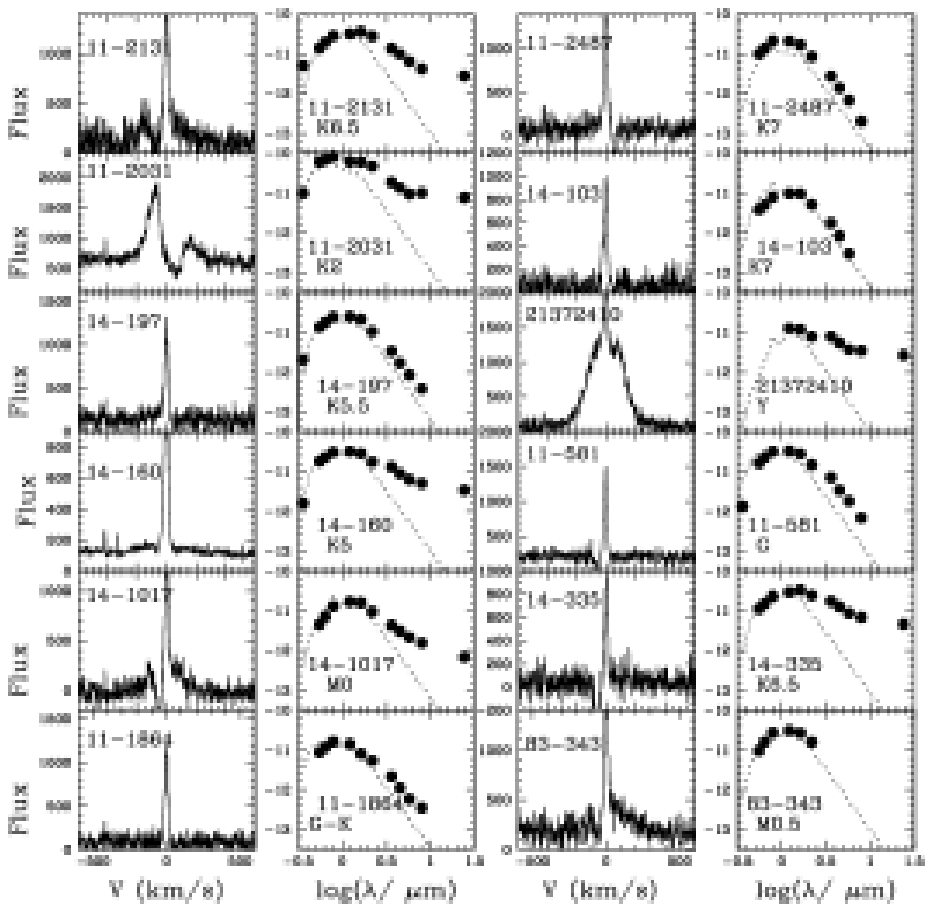}
\caption{H$\alpha$ profiles and SEDs of the observed members and potential members,
showing the optical UVRI and JHK 2MASS fluxes (Paper I, II), 
IRAC (3.6, 4.5, 5.8, 8.0 $\mu$m; Paper III) and MIPS(24$\mu$m; Paper III) data.
A photosphere \citep{kenyon95} for the given spectral type is displayed in each case
to clarify the presence of an IR excess. The 
nebular emission was successfully subtracted from the broad 
H$\alpha$ profiles with good S/N. Narrow H$\alpha$ lines, as well as some low S/N
broad profiles, are contaminated by the nebular emission.
The units of the fluxes are counts (non-calibrated) for the H$\alpha$ lines, and log($\lambda$F$_\lambda$) in
erg cm$^{-2}$ s$^{-1}$ for the SEDs (continued).
 \label{profi2}}
\epsscale{1}
\end{figure}

\begin{figure}
\epsscale{1.1}\plottwo{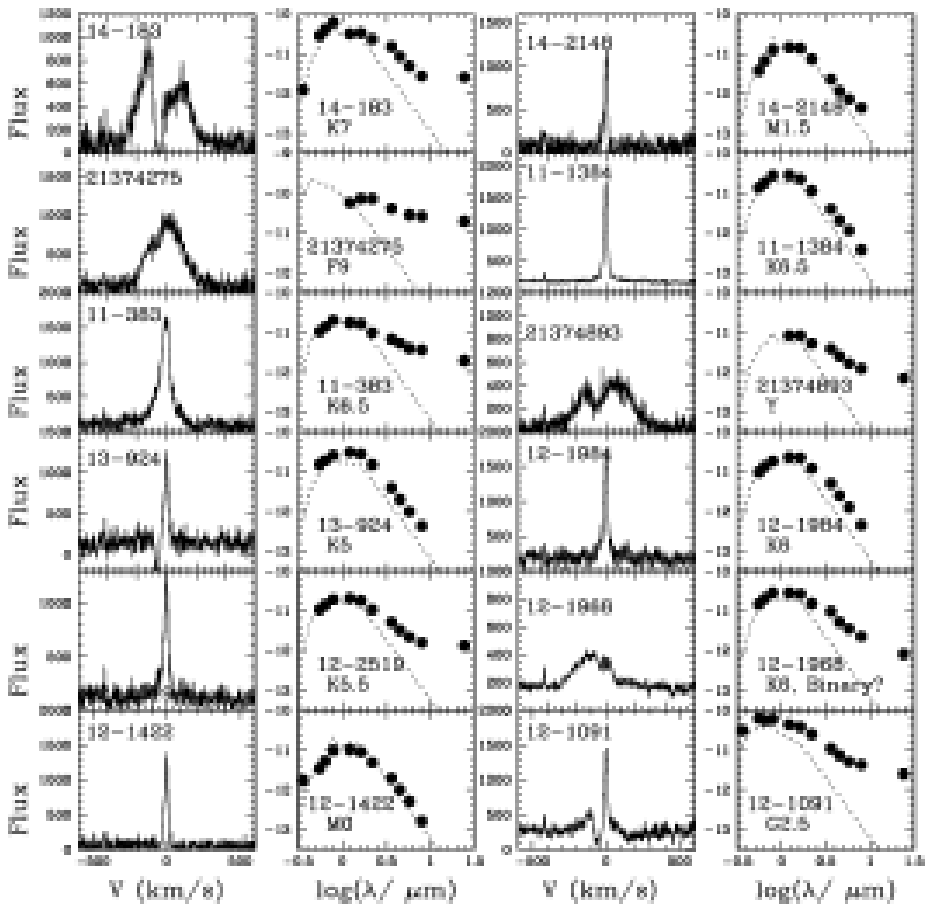}{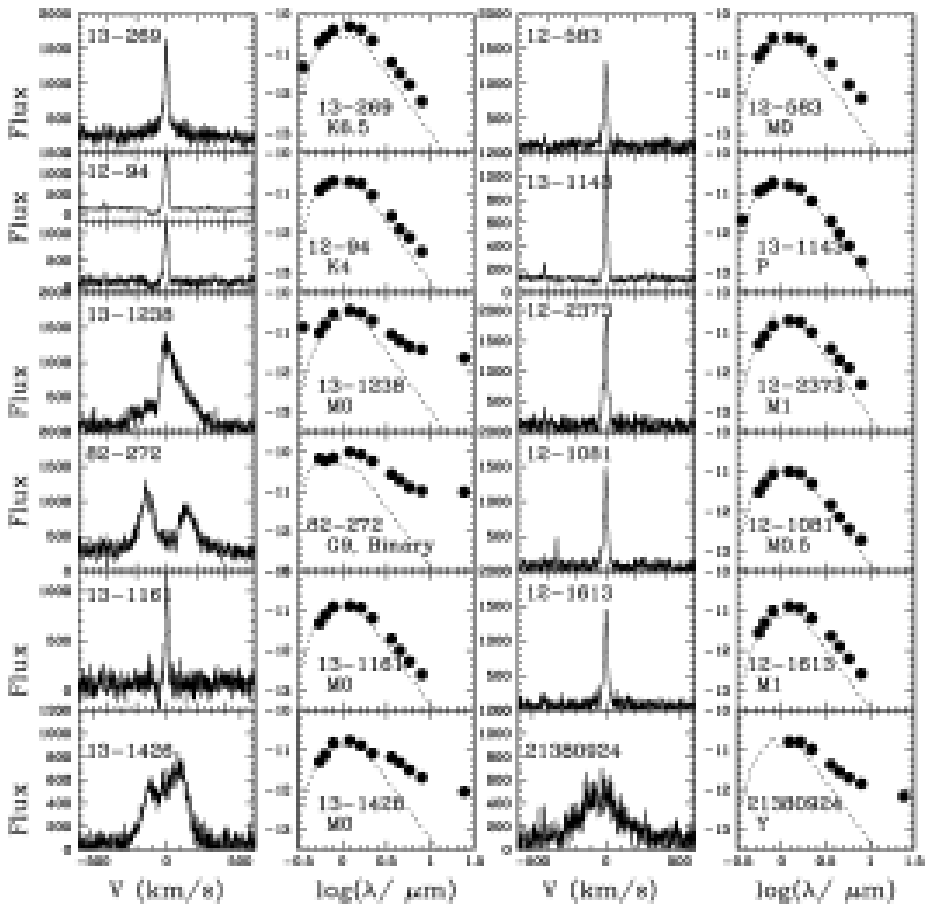}
\caption{H$\alpha$ profiles and SEDs of the observed members and potential members,
showing the optical UVRI and JHK 2MASS fluxes (Paper I, II), 
IRAC (3.6, 4.5, 5.8, 8.0 $\mu$m; Paper III) and MIPS(24$\mu$m; Paper III) data.
A photosphere \citep{kenyon95} for the given spectral type is displayed in each case
to clarify the presence of an IR excess. The 
nebular emission was successfully subtracted from the broad 
H$\alpha$ profiles with good S/N. Narrow H$\alpha$ lines, as well as some low S/N
broad profiles, are contaminated by the nebular emission.
The units of the fluxes are counts (non-calibrated) for the H$\alpha$ lines, and log($\lambda$F$_\lambda$) in
erg cm$^{-2}$ s$^{-1}$ for the SEDs (continued).
 \label{profi3}}
\epsscale{1}
\end{figure}

\begin{figure}
\epsscale{1.1}\plottwo{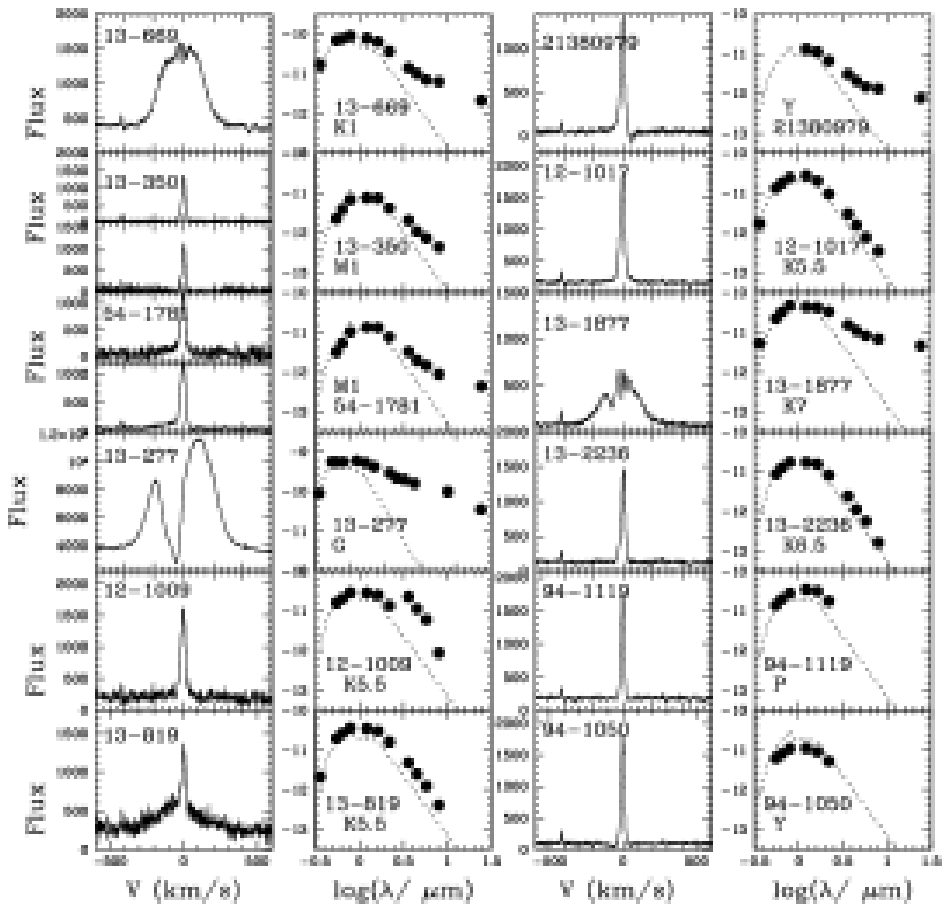}{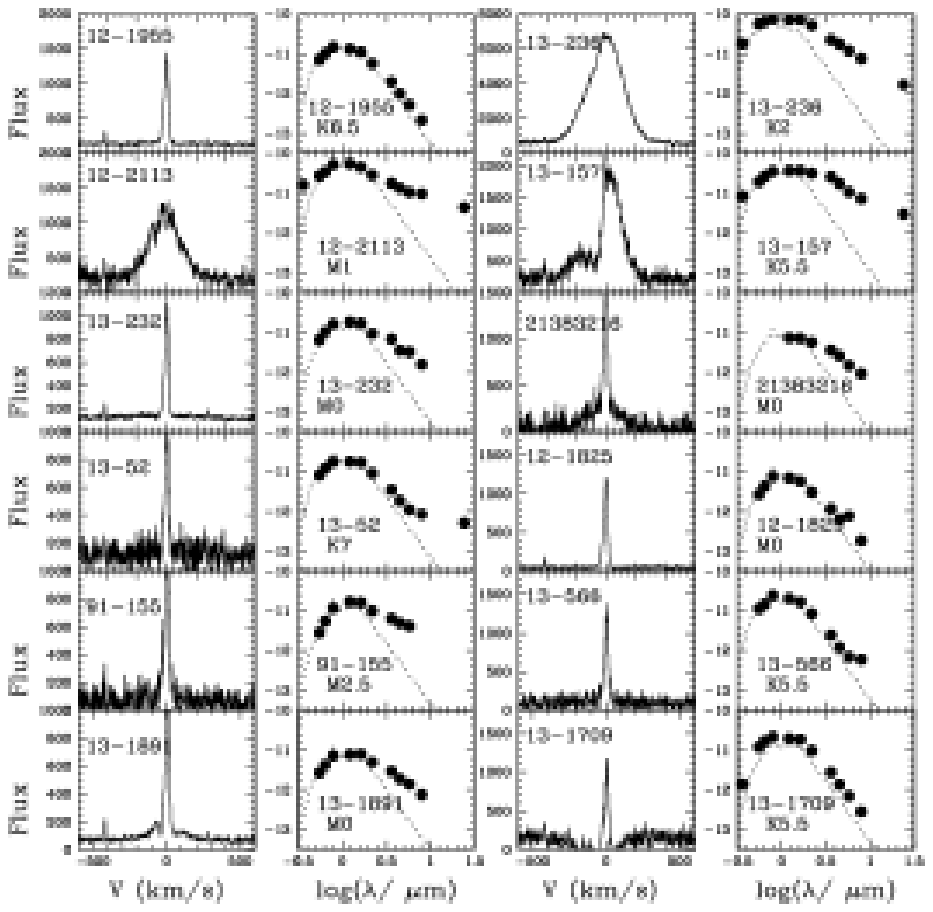}
\caption{H$\alpha$ profiles and SEDs of the observed members and potential members,
showing the optical UVRI and JHK 2MASS fluxes (Paper I, II), 
IRAC (3.6, 4.5, 5.8, 8.0 $\mu$m; Paper III) and MIPS(24$\mu$m, 70 $\mu$ only for 13-277; Paper III) data.
A photosphere \citep{kenyon95} for the given spectral type is displayed in each case
to clarify the presence of an IR excess. The 
nebular emission was successfully subtracted from the broad 
H$\alpha$ profiles with good S/N. Narrow H$\alpha$ lines, as well as some low S/N
broad profiles, are contaminated by the nebular emission.
The units of the fluxes are counts (non-calibrated) for the H$\alpha$ lines, and log($\lambda$F$_\lambda$) in
erg cm$^{-2}$ s$^{-1}$ for the SEDs (continued).
 \label{profi4}}
\epsscale{1}
\end{figure}

\begin{figure}
\epsscale{1.1}\plottwo{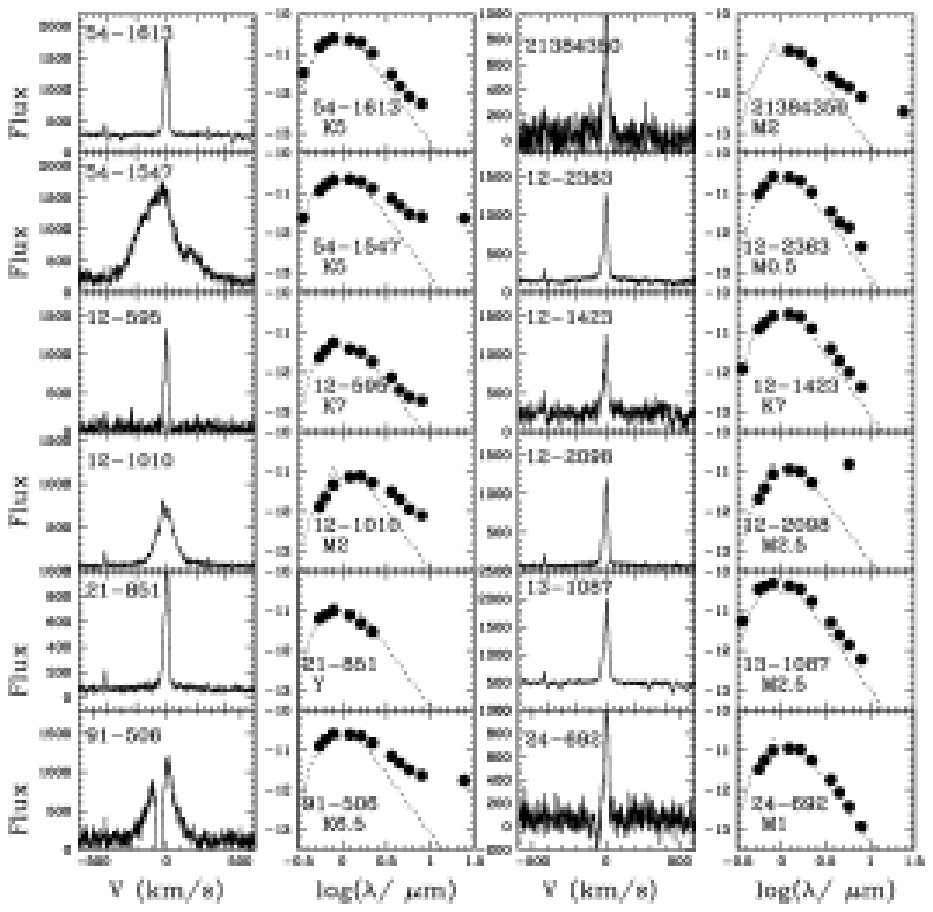}{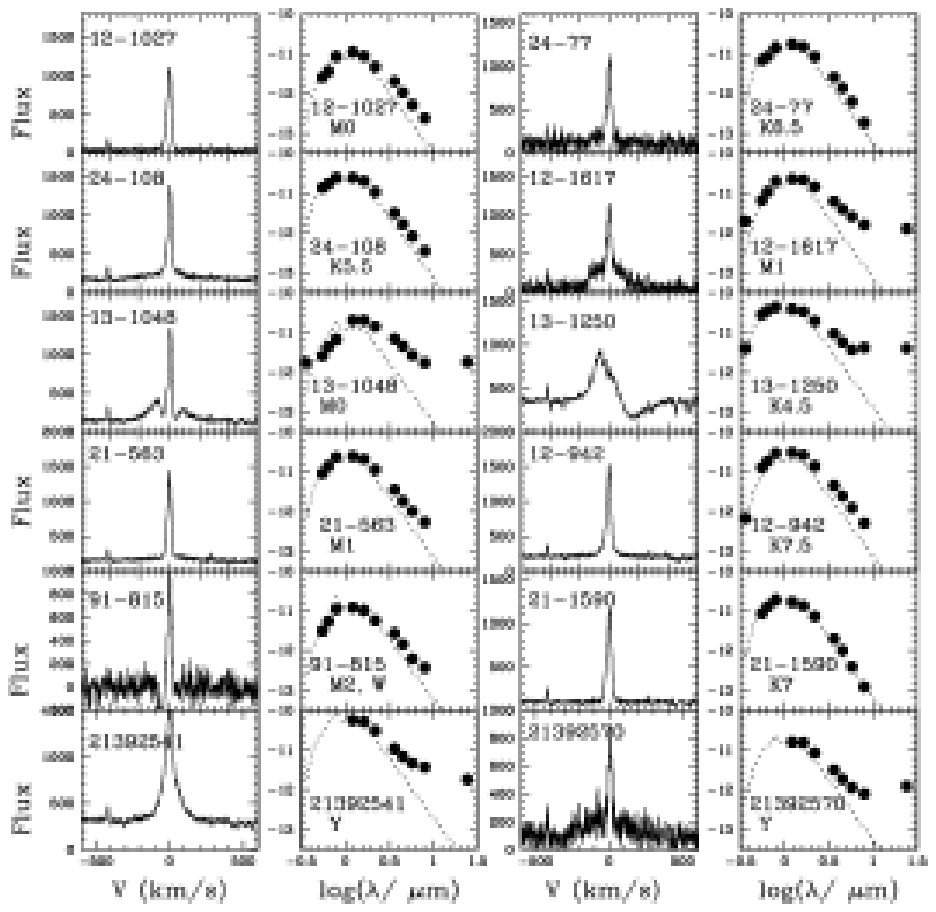}
\caption{H$\alpha$ profiles and SEDs of the observed members and potential members,
showing the optical UVRI and JHK 2MASS fluxes (Paper I, II), 
IRAC (3.6, 4.5, 5.8, 8.0 $\mu$m; Paper III) and MIPS(24$\mu$m, 70 $\mu$m for 13-277 only; Paper III) data.
A photosphere \citep{kenyon95} for the given spectral type is displayed in each case
to clarify the presence of an IR excess. The 
nebular emission was successfully subtracted from the broad 
H$\alpha$ profiles with good S/N. Narrow H$\alpha$ lines, as well as some low S/N
broad profiles, are contaminated by the nebular emission.
The units of the fluxes are counts (non-calibrated) for the H$\alpha$ lines, and log($\lambda$F$_\lambda$) in
erg cm$^{-2}$ s$^{-1}$ for the SEDs (continued).
 \label{profi5}}
\epsscale{1}
\end{figure}

\begin{figure}
\epsscale{1.1}\plottwo{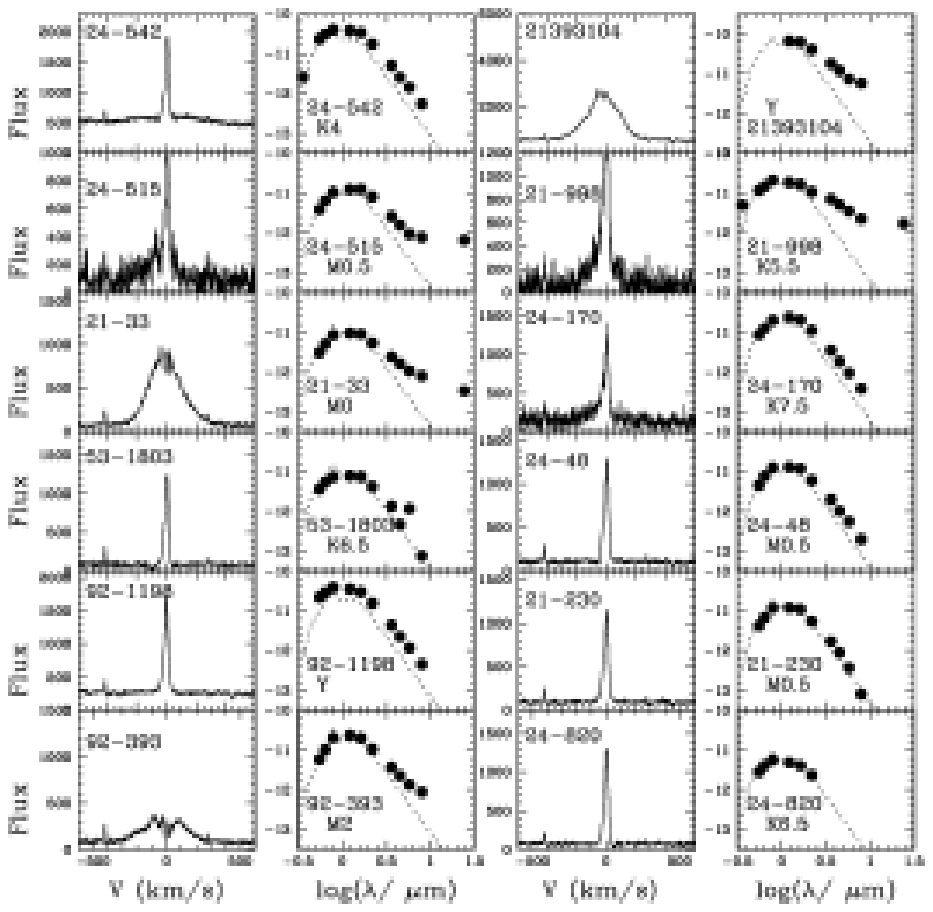}{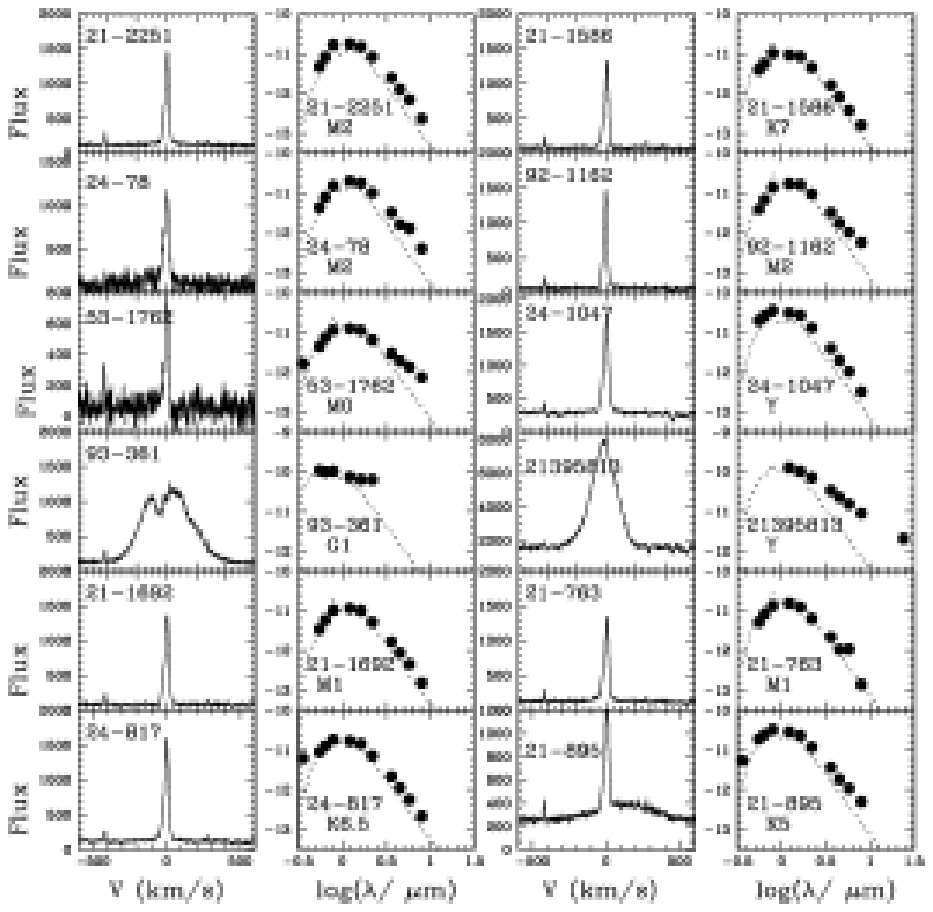}
\caption{H$\alpha$ profiles and SEDs of the observed members and potential members,
showing the optical UVRI and JHK 2MASS fluxes (Paper I, II), 
IRAC (3.6, 4.5, 5.8, 8.0 $\mu$m; Paper III) and MIPS(24$\mu$m; Paper III) data.
A photosphere \citep{kenyon95} for the given spectral type is displayed in each case
to clarify the presence of an IR excess. The 
nebular emission was successfully subtracted from the broad 
H$\alpha$ profiles with good S/N. Narrow H$\alpha$ lines, as well as some low S/N
broad profiles, are contaminated by the nebular emission.
The units of the fluxes are counts (non-calibrated) for the H$\alpha$ lines, and log($\lambda$F$_\lambda$) in
erg cm$^{-2}$ s$^{-1}$ for the SEDs (continued).
 \label{profi6}}
\epsscale{1}
\end{figure}

\begin{figure}
\epsscale{1.1}\plottwo{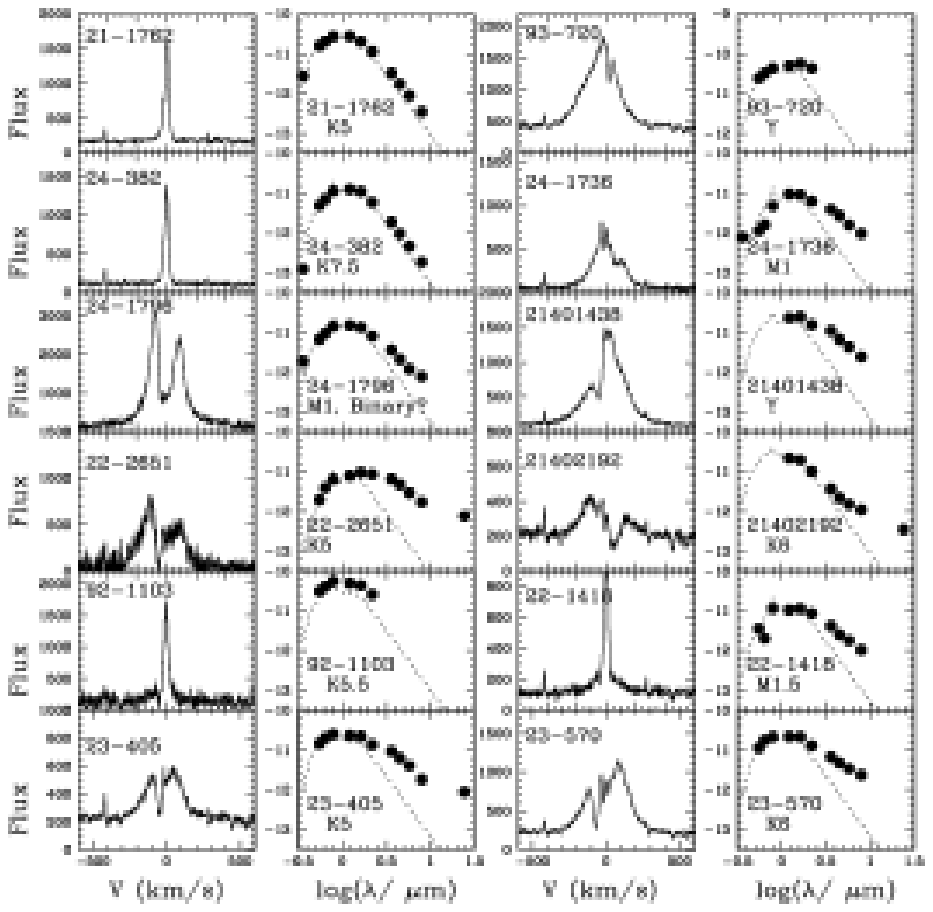}{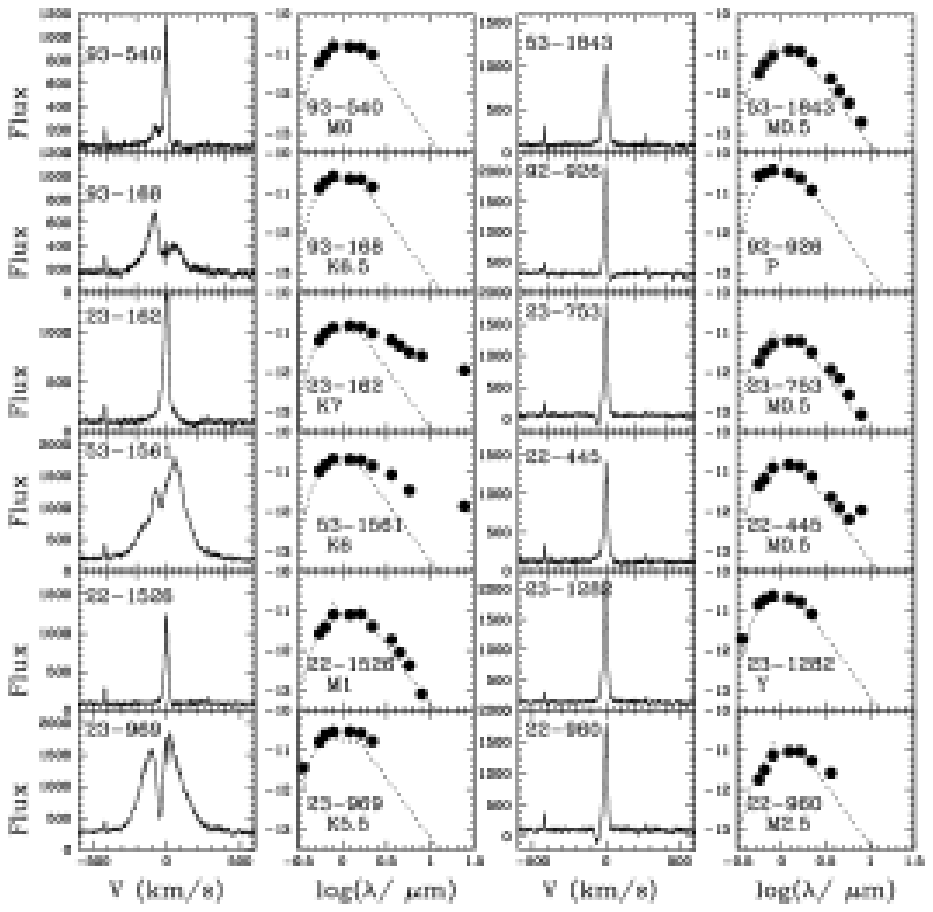}
\caption{H$\alpha$ profiles and SEDs of the observed members and potential members,
showing the optical UVRI and JHK 2MASS fluxes (Paper I, II), 
IRAC (3.6, 4.5, 5.8, 8.0 $\mu$m; Paper III) and MIPS(24$\mu$m; Paper III) data.
A photosphere \citep{kenyon95} for the given spectral type is displayed in each case
to clarify the presence of an IR excess. The 
nebular emission was successfully subtracted from the broad 
H$\alpha$ profiles with good S/N. Narrow H$\alpha$ lines, as well as some low S/N
broad profiles, are contaminated by the nebular emission.
The units of the fluxes are counts (non-calibrated) for the H$\alpha$ lines, and log($\lambda$F$_\lambda$) in
erg cm$^{-2}$ s$^{-1}$ for the SEDs (continued).
 \label{profi7}}
\epsscale{1}
\end{figure}

\begin{figure}
\epsscale{0.6}\plotone{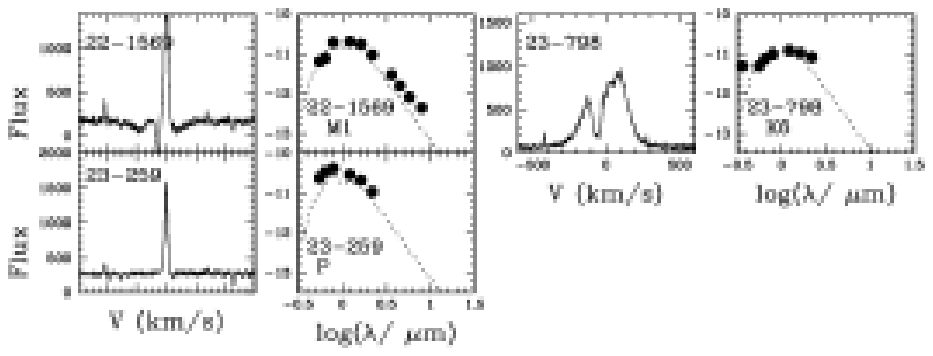}
\caption{H$\alpha$ profiles and SEDs of the observed members and potential members,
showing the optical UVRI and JHK 2MASS fluxes (Paper I, II), 
IRAC (3.6, 4.5, 5.8, 8.0 $\mu$m; Paper III) and MIPS(24$\mu$m; Paper III) data.
A photosphere \citep{kenyon95} for the given spectral type is displayed in each case
to clarify the presence of an IR excess. The 
nebular emission was successfully subtracted from the broad 
H$\alpha$ profiles with good S/N. Narrow H$\alpha$ lines, as well as some low S/N
broad profiles, are contaminated by the nebular emission.
The units of the fluxes are counts (non-calibrated) for the H$\alpha$ lines, and log($\lambda$F$_\lambda$) in
erg cm$^{-2}$ s$^{-1}$ for the SEDs (continued).
 \label{profi8}}
\epsscale{1}
\end{figure}

Figures \ref{profi1} to \ref{profi8} show the H$\alpha$ profiles of the previously
known and newly identified members and probable members, 
together with their SEDs from Paper III, that will be discussed in 
detail in Section \ref{HavsIR}. A total of 146
of the $\sim$166 Tr 37 previously known members and probable members were observed with Hectochelle,
 including 
9 objects with uncertain membership. Membership is revised in Section \ref{velocities},
where we also analyze the potential members selected among optical and infrared
candidates.
Table \ref{spec-dat} lists the previous and new spectroscopic information, the H$\alpha$ EW 
and the full width measured at 10\% intensity (when the nebular
emission was successfully subtracted), and the membership 
(based on this and on our previous studies).

\subsection{H$\alpha$ emission: line profiles, EW, and accretion \label{haprofi}}

The presence of active magnetospheric accretion can be inferred observing
the emission lines in young stars, especially the H$\alpha$ emission. 
The large velocities of the
material involved in the magnetospheric
accretion processes produce strong emission lines
with broad velocity wings, usually larger than 
$\pm$  100 $\kms$ \citep{reipurth96,hartmann98,muzerolle98a,muzerolle01,white03,bonnell98}.
In general, accretion rates over 10$^{-12}$ M$_\odot$/yr result in broad H$\alpha$
velocity wings \citep{muzerolle03}, compared to the 10$^{-9}$-10$^{-10}$ M$_\odot$/yr
limit imposed by the detection of U band excesses (see Paper I,II and Section \ref{accretion}).
H$\alpha$ emission is produced as well in the chromosphere of young,
non-accreting stars, but in this case, the line is narrow
and does not present high velocity wings \citep{appenzeller89}. 
Therefore, the distinction between accreting and non-accreting stars (defining
``non-accreting'' as showing H$\alpha$ emission similar to
the diskless WTTS or Class III objects; see Section \ref{accretion} for details) 
becomes clear from high-resolution spectra.

Examining the H$\alpha$ profiles, we can observe the different
characteristics related to accretion \citep{reipurth96,muzerolle98a,muzerolle01}: high-velocity
wings revealing accretion velocities that can be sometimes
larger than 300 km/s (see the profiles of 11-2037, 13-277, 93-720
and 93-261 among others),
blue-shifted and red-shifted absorption features (11-2322, 12-1091, 13-277, 13-1048, 21402192+5730054, 
among others), 
and inverse P-Cygni profiles
characteristic of low accretion rates (11-2031, 13-1250). 
The most common profile is characterized by a blue-shifted absorption, although red-shifted
absorption and absorption with no velocity shift are also observed, as well
as several symmetric profiles, and 3 cases of inverse P-Cygni profiles
(always related to objects with very low accretion rates).
The characteristics of broad H$\alpha$ profiles are listed in Table 2.
Comparing with the spectra obtained for the younger ($\sim$ 1 Myr)
ONC stars \citep{sic05onc}, the profiles are similar, but
it is worth to note the lower number of very
strong emission line stars with large velocity wings in the 4 Myr-old Tr 37,
as well as the smaller H$\alpha$ EW of the broad-lined stars.
Given that the spectral types of the stars range from mid-G to M2 in both samples, being
roughly comparable,
the observed differences should be related mainly to accretion and
disk evolution. 

It is interesting to note the cases of stars like 73-311, 21362507+5727502,
11-2131, 14-1017 and 13-819, among others, for which the broadening 
of the H$\alpha$ profiles is relatively small, showing velocities around 
only $\pm$100 km/s, but visibly larger than for
non accreting stars (Muzerolle et al. 2003; Natta et al. 2004). Their EW
are in the lower limit for distinguishing CTTS from WTTS or even below, so they would
be classified as WTTS attending to the EW \citep{white03}. 
Low H$\alpha$ EW can be also found among some of the stars showing inverse
P-Cygni profiles, characteristic of low accretion rates 
\citep{muzerolle98a, muzerolle98b}.
This is the case of 21402192+57300054 (EW = -4 \AA\ \footnote{Negative values of the EW stand
for emission.} , spectral type K6)
and 13-1250 (EW = -4 \AA\ , spectral type K4.5).
The small EW in these broad-H$\alpha$ (and thus, accreting)
stars accounts for the apparent contradictions between the presence of U and
IR excesses and the measurement of a ``weak'' H$\alpha$ EW 
from low-resolution spectroscopy (Paper I, II; we will discuss 
the relation between IR excesses and accretion in
Section \ref{HavsIR} in more detail). 
To summarize, the EW values for accreting stars range from -2 to -80 \AA\
approximately, with an average
EW is -27\AA\, with a median value of -20\AA\ and a standard
deviation of 19\AA\  (due to the large differences in EW
present in the sample).

With the high-resolution spectra, we are also able to detect the presence of an
accreting spectroscopic binary with double lines (SB2), 82-272, with EW = -7 and -8 \AA\ for each
of the components, respectively. The double-peaked profile of 24-1796 could be
indicative of another accreting binary, even though given the smaller offset of the
two peaks, the binarity is not so evident in this case.
Finally, the wavelength offset for the broad H$\alpha$ in 12-1968 
may suggest a binary containing an accreting and a non-accreting component.
No other SB2 are evident within this sample.
We also want to mention the broad blueshifted absorption seen in the
two spectra taken for 12-94, a K4 weak-lined star whose membership is uncertain
(from both Li detection and radial velocity).
Finally, there are some (mostly low S/N) objects, for
which the width of the H$\alpha$ line is not clear. These objects
are marked with the uncertainty flag in Table \ref{spec-dat}, and due to the
absence of near-IR excess and other signs of accretion
in most of them (see Section \ref{HavsIR}), we presume they are likely to 
be narrow-lined, non-accreting WTTS. Followup of these
objects would be highly recommendable, especially in the case of 73-758, which 
has a SED typical of a ``transition object'' (see
Section \ref{transition} for a more detailed description).

\subsection{Membership and dynamics\label{velocities}}

As we mentioned previously, the membership (or probable membership) of 146 objects
was determined previously using low-resolution spectra (Paper I, Paper II)
to detect the presence of Li 6707\AA\ absorption, the presence of
H$\alpha$ emission (both characteristic of young stars), and the
spectral type and extinction (found to be consistent with the cluster).
The broad H$\alpha$ profiles in the Hectochelle data is another proof
of membership that allows us to confirm the accreting known members,
and to identify 7 new ones. Due to the
presence of unsubtracted nebular H$\alpha$, we cannot
establish the membership of the stars with narrow H$\alpha$ emission 
via their H$\alpha$ emission only (this
would be otherwise a criterion for detecting new WTTS members, since
H$\alpha$ chromospheric emission is another sign of youth, see Hartmann 1998
among others).
Nevertheless, comparing the radial velocities of the newly observed members to the
radial velocities of the previously known members and stars with
broad H$\alpha$ lines, we can
determine the membership of the rest of newly observed objects, and
check the cases of dubious membership.

\begin{figure}
\plotone{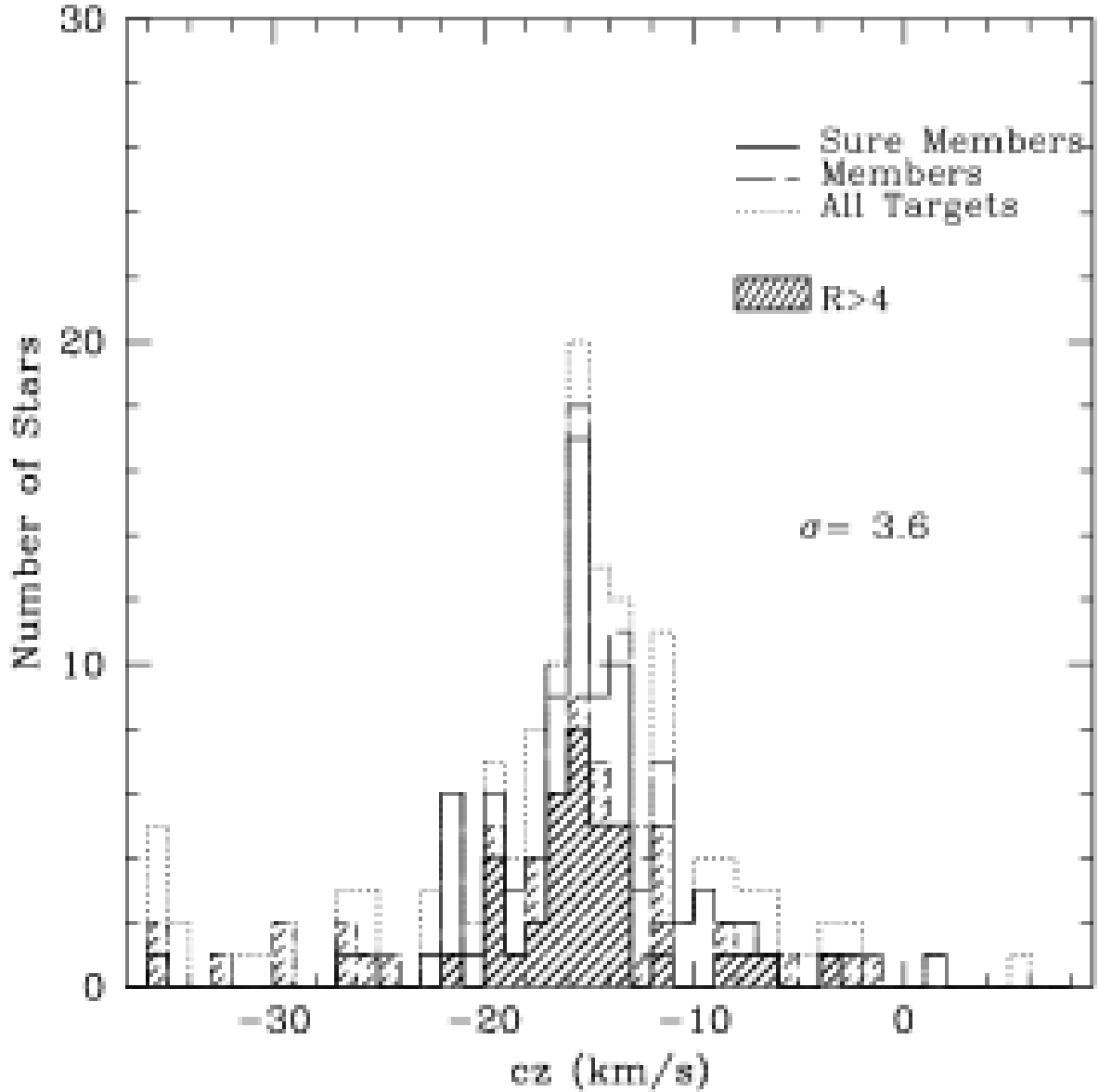}
\caption{Radial velocity histogram for Tr 37. The radial velocities of
sure members (confirmed by Li 6707\AA\ in low-resolution spectra and/or
broad H$\alpha$ lines, see text), sure and probable members (labeled 
``members''), and the rest of observed stars (``all targets'') are
displayed in a histogram. Only velocities obtained through good cross-correlations
are considered. Shaded areas represent the stars with R$>$4. Using the 
sure members with R$>$4, we obtain an average radial velocity for the
cluster cz =  -15.0 $\pm$ 3.6 km/s, where $\sigma$ = 3.6 km/s
is the standard deviation.
 \label{histo-radvel} }
\end{figure}

Figure \ref{histo-radvel}
shows the radial velocity histogram of the cluster members. Taking into account
only the ``sure'' members (with membership confirmed via clear Li 6707 \AA\ absorption
and/or by the presence of broad H$\alpha$ emission in the high-resolution spectra), and
considering only the high S/N cross-correlations (R$>$4; 37 stars in total), we can estimate the
average cluster radial velocity to be cz = -15.0 $\pm$ 3.6 km/s, where the velocity
dispersion is $\Delta$V= 3.6 km/s. Note that for this calculation 
we excluded the stars with large offsets in Figure \ref{histo-radvel}, as they
are likely to be single-line spectroscopic binaries (SB1). This velocity 
dispersion accounts in part for the real expansion velocity of the cluster, but it is
largely affected by the individual errors (which are typically 1-2 km/s) and by the $\sim$0.6-0.8 km/s
extra error mentioned in Section \ref{observations}). Taking into account the known members and the
total number of observed stars, we can estimate the probability of being a member if
the radial velocity of the object falls within a certain number of sigmas from the
cluster average radial velocity. If we assume equal probabilities for a
random star to have a certain radial velocity, counting the number of stars observed and
the number of known members, we could estimate the ``average number of stars'' that
would fall randomly within the 1,2 and 3 sigma bins, and compare it to the number of
known members. This way, we can estimate the stars within the 1$\times \Delta$V bin to
have a probability of around $\sim$90\% of being cluster members, a probability of being
member of $\sim$75\% if the star falls within 1-2$\times \Delta$V, and a probability of $\sim$40\%
of being member if the star falls within the 2-3$\times \Delta$V bin. These are only approximate
values, and membership may be more uncertain for the stars with low R,
so we define the stars within 1$\times \Delta$V to be ``sure members'' (named `Y'
in Table \ref{spec-dat}), the stars within
2-3$\times \Delta$V to be ``probably members'' (`P'), and the stars within 
2-3$\times \Delta$V to be ``probably non-members''(`PN'). 
We give in any case priority to the membership established from Li absorption, 
from the presence of clear IR/Spitzer excess, or from the 
broad H$\alpha$ emission from the high-resolution spectra. The stars with 
certain membership, good correlations and high non-cluster radial velocities 
are likely to be single-line binaries (SB1) or probable
SB1 (when the cross correlation has large errors and/or the radial velocity is off but very close to the cluster
velocity), see Table \ref{spec-dat}.

With this procedure, we have identified a total
of 14 new sure members (labeled `Y'), 9 probable members (`P'), and 9 probable non-members or
lower-probability candidates (`PN'). Among these new members,
5 could be confirmed based on their strong H$\alpha$ emission
(2 of them being sure members, 2 probable members, and 1 low-probability
member). Using radial velocities, we have revised the 
stars with dubious membership. These
stars had low S/N low-resolution spectra, which did not allow us
to determine clearly the presence of Li 6707\AA\ absorption, or
are G stars, for which the smaller Li EW  was under our detection limit
(Paper I, II). Radial velocities confirm the membership of 
13-350, 13-1891 and 24-820. On the other hand, 11-581 and 12-94
show non-cluster radial velocities, being probably non-members or SB1 (we
reject them as members in order to be conservative).
For the rest of dubious members (14-306, 11-1067, 12-583 and 12-1613),
the S/N in the high-resolution spectra does not let us obtain accurate
radial velocities, so we cannot confirm nor reject them.
Summarizing, and naming as safe all objects with strong H$\alpha$, 
Li detection (in low-resolution spectra) and/or radial velocities
within the 1$\sigma$ region (`Y'), we have 157 sure members, 11 probable
members, 8 low-probability members, and 2 non-members (rejected from the
initial sample in Paper II).

Taking into account the errors in the radial velocity dispersion,
we are approximately at the edge of being able to detect the true
velocity dispersion or expansion of the cluster. Our main limitation
is the poor S/N resulting in large errors in the individual radial velocities, which
are up to 1-2 km/s even in the cases with R$>$4.  Considering the
estimated mass of the cluster from the estimated number of cluster
members ($\sim$300 members, or M$\sim$300 M$\odot$ in the bulk of the
cluster, see Paper III), and its size ($\sim$4 pc radius for the bulk of the population, Paper II),
it is most likely unbounded and expanding \citep{spitzer84}.  
The radius of the bulk (4 pc) is consistent with the expected size
after 4 Myr, assuming a typical velocity dispersion of $\sim$1 km/s.
Nevertheless, the Tr 37 region is likely to have triggered at least one population,
aged about 1 Myr, in the globule (at distances 4-5 pc from the cluster center; Paper II, III),
and maybe another associated population about 5-6 pc North (Paper II, III),
while expanding in a non-uniform way. Given the precision of our
radial velocities, and the reduced number of stars in these
two populations, we are unable to detect any difference in their radial velocities
at this point.

\subsection{Accretion versus IR excess \label{HavsIR}}

According to Figures \ref{profi1} to \ref{profi8}, the agreement between the presence of 
broad H$\alpha$ emission and the detection of an IR excess is remarkable.
It can be also noted that the brighter disks (stronger IR 
excesses with respect to photospheric levels at all wavelengths, resembling
younger disks) and the SEDs
with strong U excesses are normally associated to the strongest and
broadest H$\alpha$ emission lines (see 13-277, 12-236, 13-1877
among others), as we would expect if the evolution of the accretion
processes proceeds in parallel to the evolution of the dusty disk.
In general, the high-resolution H$\alpha$ spectra 
solve the apparent contradiction
between small H$\alpha$ EW (according to White \& Basri 2003, 
WTTS have EW $<$ 10\AA\ in emission for spectral types K6-M2, and EW $<$ 3\AA\
in emission for spectral
types K5-G) and the presence of an IR excess that we found in
Paper II and Paper III. Indeed, the puzzling stars from Paper III
have all very small H$\alpha$ EW, so they would be classified as WTTS
according to White \& Basri (Paper III), 
if the high velocity wings were not resolved. 
These are the objects 11-1209 (spectral type K6, EW = -6/-4 \AA\ from high-
and low-resolution spectra, respectively), 12-1968 (K6, EW = -8/-11 \AA),
13-819 (K5.5, EW = -6/-10 \AA), 13-1048 (M0, EW = -8/-7 \AA),
13-1250 (K4.5, EW = -4/-2 \AA), 21402192+5730054 (K6, EW = -4/-8 \AA), 
and 93-540 (M0, EW = -5/-18 \AA). The small EW suggest low accretion
rates, as we will present in Section \ref{accretion}.

\begin{figure}
\plotone{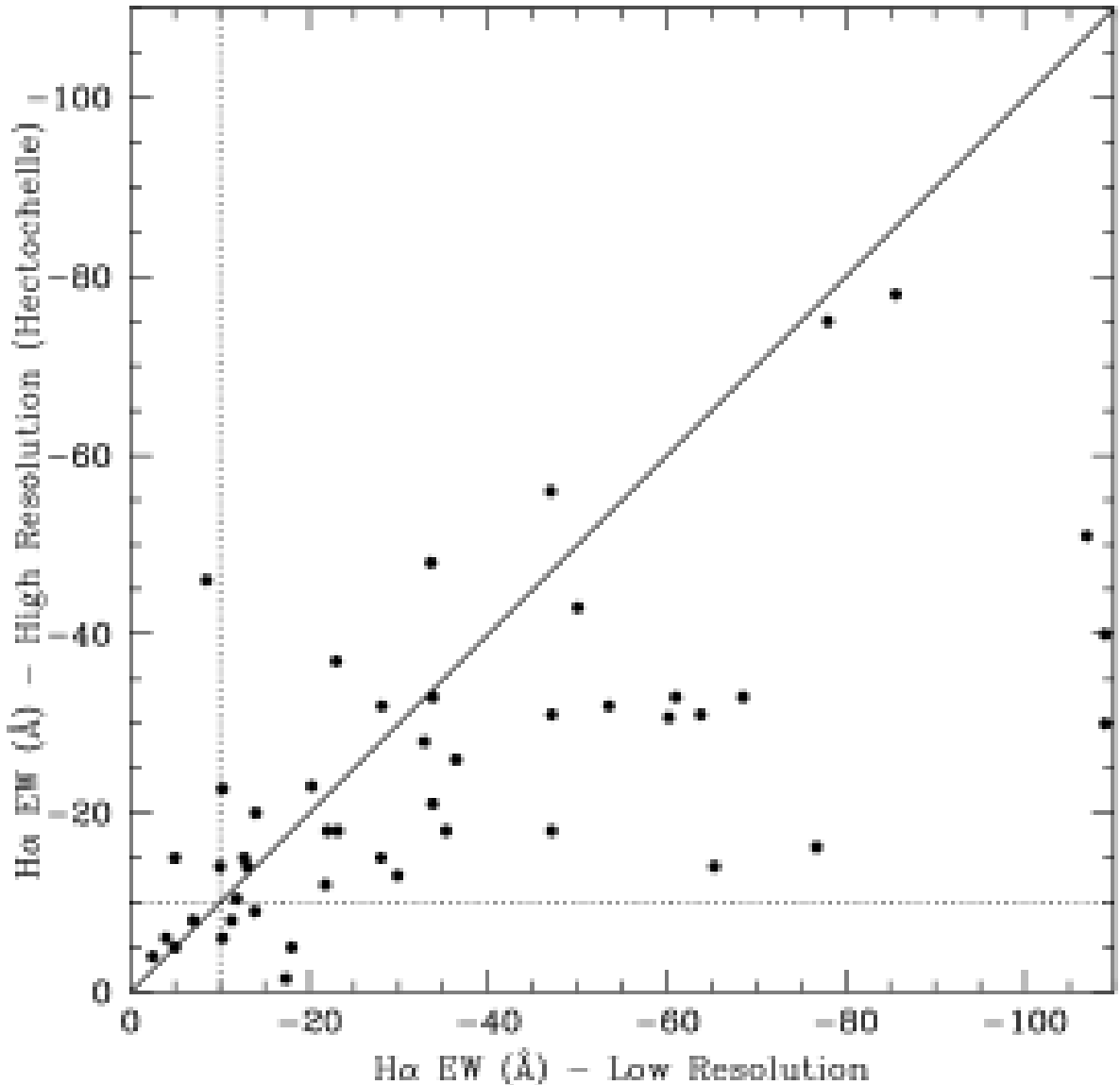}
\caption{Comparison of the H$\alpha$ EW (in \AA) measured using low resolution
(Hectospec and Hydra, see Paper I, II) and high resolution (Hectochelle) spectra.
Note that we were able to measure the EW only for spectra with broad H$\alpha$ 
lines. Therefore, all objects displayed are accreting stars or CTTS.
The dotted lines indicate EW=-10\AA\, which is the classical limit 
for distinguishing CTTS and WTTS, although we find several broad-lined
stars with smaller EW. The dispersion between the H$\alpha$ values measured
with low and high resolution is mostly due to the intrinsic variations of the
stars \citep{sic05onc}. Some extreme values of H$\alpha$ measured from 
low-resolution spectra
are caused by poor S/N, which produces very low continuum in the region.
Note that some variation can be introduced because of poor continuum
estimates in cases of low S/N (see Paper I, Paper II for more details),
resulting in large ($\sim$60-100 \AA) EW that are not real.
 \label{hacomp} }
\end{figure}

There is a good agreement between the measured H$\alpha$ EW
from high- and low-resolution spectra (Figure \ref{hacomp}).
The observed variations of the H$\alpha$ EW are consistent with the
typical variability observed in the ONC (Sicilia-Aguilar et al. 2005;
note that some very large EW$\sim$-60 to -100\AA\
found in low-resolution spectra are due to low signal-to-noise
resulting in a poor continuum estimate). The CTTS with small EW
represent a small fraction of the total sample, suggesting that the
classification of White \& Basri (2003) successfully distinguishes
accreting (CTTS) and non-accreting (WTTS) stars in more than 90\% of the cases.
Therefore, an H$\alpha$ EW survey is probably good enough to determine
accretion in very young regions, which contain a large number of accreting
stars with relatively large accretion rates and large EW. In older regions (like Tr 37),
where the accretion fraction and the accretion rates are much smaller, 
determining accretion by H$\alpha$ EW alone is uncertain, 
so it should be combined with high-resolution
spectroscopy and/or near- and mid-IR observations.

Despite the general good agreement of IR excesses with 
high-velocity wings in the H$\alpha$ line, there are a few
cases for which the conflict could not be resolved. Some of these 
cases are the ``transition objects'' (see Paper III), which we
discuss extensively in Section \ref{transition}, since about half of 
them, despite having (outer, optically thick) disks, are most likely not accreting. 
Two other ``contradictory'' objects have apparently normal IR excesses and no 
strong H$\alpha$ velocity wings. Due to the discrepancies
between the high- and the low-resolution spectra 
taken in former campaigns, and due to the fact that the spectra are
very noisy and weak, we believe that an small position offset in
the Hectochelle fibers may have occurred in some of the
cases. These offsets, although unlikely, may occur in a couple of
fibers in each Hectochelle setup, so the low number of conflicting
spectra is consistent with the offset hypothesis.
These stars with contradictory low- and high-resolution spectra are 
21364596+5729339 (classified as a class I object, no spectral type
given due to heavy veiling, H$\beta$ in emission was clearly detected, as well as some
broadening of H$\alpha$ even at 5 \AA\ resolution) and 
14-335 (EW = -18 \AA\ in the low-resolution spectra, H$\beta$
in emission and Li 6707 \AA\ in absorption were clearly detected with Hectospec).
An offset occurred during the Hectochelle run would explain
the extremely low S/N of 14-335 (we would be observing mostly the background), 
and the relatively luminous Hectochelle spectrum in 21364596+5729339 would 
mostly contain
the emission lines from the globule in the surroundings of the star.

The stars 14-160 (spectral type K5) and 23-162 (K7)
show strong H$\alpha$ with very small velocity wings,
which resemble the spectra of WTTS, although the intensity of the 
H$\alpha$ emission is much higher than the emission of the normal WTTS and the nebula.
They show very strong but relatively narrow H$\alpha$ emission,
with no velocity wings extended beyond $\pm$50-80 km/s. Their low-resolution
spectra present H$\beta$ and H$\gamma$, and H$\alpha$ EW = -22 \AA\ and -7 \AA\ for
14-160 and 23-162, respectively. There is no evidence of contamination nor large errors
in their IRAC/MIPS photometry, and the nearby spectra show weak nebular
emission that cannot account for the strong and narrow H$\alpha$ observed in these 
two stars. The small EW and the lack of U band
excesses suggest very small accretion rates. A configuration where
the H$\alpha$ profile appears narrow despite accretion is rare,
requiring a special geometry and/or view angle, and a small accretion rate
(at least 10$^{-10}$ M$_\odot$/yr or less, see Muzerolle et al. 1998a, 2003).
Nevertheless, our sample is large enough to be likely to contain some
of these rare profiles.
Note that accretion rates under 10$^{-12}$ M$_\odot$/yr 
would not produce line broadening, although given the ``normal'' disk emission of these
two stars, such low rates would be rare (Muzerolle et al. 2003).

Less than 5\% of the objects (the stars 91-155, 12-2519, 22-1418, 53-176 and 54-1781) 
show an apparent contradiction between the
presence of narrow H$\alpha$ and a normal-looking disk. Their low-resolution spectra
show H$\beta$ emission for 91-155 and 12-2519, even though the H$\alpha$
EW typical of WTTS, which would indicate no accretion. 
Contamination by a nearby star is likely for the star 91-155,
which shows an offset in its IR data with respect to the optical information
(our IRS spectra seem to confirm the presence of a disk that may belong to the
companion, Sicilia-Aguilar et al. in preparation).
For all the other cases, the presence of nearby stars and/or small patches of nebulosity
makes uncertain the IRAC photometry.
The stars 12-1009 and 13-232 show intrinsic large errors in their IRAC photometry,
so the presence of a disk is highly uncertain (Paper III).

We only find 1 object with no IR excess
and some broad H$\alpha$ emission,  13-819. This K5.5 star does not have
any excess up to 8$\mu$m, but since it is relatively faint,
an excess at 24 $\mu$m of less than $\sim$20 times the photospheric level
would be undetectable with our MIPS photometry. The lack of
8$\mu$m excess suggests that the silicate feature is very faint or absent (as the 8$\mu$m
IRAC band covers nearly up to 10$\mu$m), suggesting a strong depletion of
small grains at distances close to the star, and a gap $>$5 AU.
Its small H$\alpha$ EW is consistent with the very small accretion rate detected with U band 
(\.{M}=4$\times$10$^{-10}$-2$\times$10$^{-9}$ M$_\odot$/yr). Given that the 24 $\mu$m band traces
material at distances around 10-20 AU (depending on spectral type),
this would suggest that accreting stars with inner gaps of 10-20 AU
are extremely rare (around 1\% of the total number of disks, $\sim$0.5\% of the
total number of members), have very short lifetimes (about ten times shorter than
the typical lifetimes for ``transition objects'' with few-AU gaps), 
and/or have extremely weak accretion
rates (undetectable in H$\alpha$, \.{M}$<$10$^{-12}$ M$_\odot$/yr). 
Observations at longer wavelengths would
be an interesting continuation of this work, not only to detect stars with 
large gaps, but to follow the evolution (in mass and size) 
of the outer disk in these stars
where the inner disk is considerably evolved.

To summarize, the Hectochelle high-resolution spectra can explain the apparent
contradictions found between the H$\alpha$ EW and the presence of disks
inferred from IR excesses. With the Hectochelle data, we can consistently
explain 95\% of the objects observed, with the remaining
5\% corresponding mostly to stars with poor photometric and/or
spectroscopic data.

\section{Discussion \label{discussion}}

\subsection{Accreting and non-accreting ``transition objects'' \label{transition}}

One of the most interesting type of objects for the study of the processes leading
to disk dissipation and planet formation are the ``transition objects''.
We define as ``transition objects'' the stars showing 
IR excess only at the longer wavelengths (typically, from the IRAC 5.8$\mu$m band on; Paper III).
Normally, these stars exhibit H$\alpha$ EW (from low-resolution spectra)
consistent with WTTS, and the discrepancy cannot be explained in terms of contamination by nearby
stars, the emission from the globule (which can be important at 5.8, 8 and 24 $\mu$m),
nor because of photometric errors (Section \ref{HavsIR}).
The lack of IR excess at shorter wavelengths 
suggests that important dust settling and/or grain growth have occurred in the
innermost part of these disks. Since accretion indicators
point to very low or even no accretion, the presence of larger bodies (planets?) in the
inner part of the disk, which might be responsible for the absence of
gas flow (Lin \& Papaloizou 1986; Bryden et al. 1999; D'Angelo et al. 2003; Quillen et al. 2004), 
cannot be ruled out. Planets have been invoked as a mechanism to 
clear up a gap in the disk, preventing in some cases accretion, as the inner disk
cannot be replenished with gas coming from the outer disk (Forrest et al. 2004;
D'Alessio et al. 2005). There are well-known cases of stars with non-detectable
near-IR excesses and accretion (GM Tau and TW Hya are perhaps the most famous,
see Calvet et al. 2002, Uchida et al. 2004) and with no accretion (CoKu Tau/4; Forrest et al.
2004). The lack of significant U band excess suggests
very low accretion rates ($\sim10^{-9}-10^{-10}$ $M_{\odot}$/yr),
so of high-resolution spectra are required in order to clarify the nature
and presence of gas accretion in these objects.

Using Hectochelle in Tr 37, we can confirm the presence of both accreting and non-accreting
``transition objects'' (see Section \ref{accretion} for a more detailed description of 
accretion rate limits in objects with no velocity wings). 
In Paper III, our definition of ``transition objects'' emphasized these
objects with inner gaps and no accretion (small H$\alpha$ EW and no U excess).
Given that Hectochelle has revealed the presence of both accreting and non-accreting
``transition objects'', here we modify our definition to include all objects
lacking near-IR excesses, independently of their accretion status. We have also
revised the Spitzer photometry for this study, and although there are no
significant changes, we have dropped from our list objects likely to be
contaminated by either emission from the globule and/or nearby stars (these
are the cases of 73-184 and 24-170). From the ``transition object'' list in Paper III, we 
did not observe two potential ``transition objects'', 12-705 and 22-1569,
which should be kept in mind for completitude and are good candidates for
further study. Otherwise, the list stated here is the most complete and revised
to the date. Note that, in any case, the number of ``transition objects'' is
likely to be wavelength-dependent (Hartmann et al. 2005), 
and given that our data extend only to
24 $\mu$m, we are not sensitive to objects with inner holes larger than a few
AU, unless we are able to detect active accretion.

The stars 14-11, 13-52, and 13-566
are clearly not accreting, showing no velocity wings and profiles typical of
WTTS despite of their large IR excess at wavelengths
longer than 5.8 or 8 $\mu$m. The stars 73-758, 21384350+5727270, 12-595, and 21392570+5729455 are
most likely non-accreting as well, even though their 
spectra are noisy and might have some undetected higher velocity wings.
On the other hand, the stars 13-1250 (K4.5, EW = -4/-2 \AA), 
21392541+5733202 (spectral type unknown, EW non-measurable), 
24-515 (M0.5, EW non-measurable), 92-393 (M2, EW = -21/-34 \AA), 
21402192+5730054 (K6, EW = -4/-8 \AA) and 13-819 (K5.5, EW= -6/-10 \AA) show H$\alpha$ profiles characteristic 
of accretion, in spite of their lack of near-IR excess. 
It is worth to mention that all the accreting
``transition objects'' (except for 92-393\footnote{This star has an spectral type M2. According to
White \& Basri 2003, stars of spectral types later than M2 may have chromospheric
values of H$\alpha$ up to 20\AA, so this would place 92-393 in the limit between CTTS and
WTTS as well}) have very small H$\alpha$ EW, comparable to WTTS, 
or in the limit between CTTS and WTTS
(non-measurable in the cases of 21392541+5733202 and 24-515 due to difficulties subtracting the
background component when the velocity wings are so small). The small EW is suggestive of a very
small accretion rate, what we can corroborate based on the U band observations presented in
Paper II. The only stars with detectable U band excess are 13-1250 and 13-819, which were observed in
two different campaigns. For 13-1250 (which may also have a very small excess over
the photosphere), one of the observations
results in non-significant U band excess, and the other observation reveals a small excess
(in the limit of detectability) that could be related to an accretion rate 
around 10$^{-9}$ $M_{\odot}$/yr. For 13-819, the accretion rate varies from
4$\times$10$^{-10}$ to 2$\times$10$^{-9}$ M$_\odot$/yr, confirming
our predictions of very low accretion rates.

Taking into account the number of bona fide 
``transition objects'' ($\sim$10\% of the stars with disks), 
and comparing to the number of disks expected to have disappeared 
within the past 3 Myr (considering that regions aged $\sim$1 Myr 
have $\sim$ 80\% of disks, and that the 4 Myr-old Tr 37
has only $\sim$45\% disk fraction, and assuming a constant rate of disk
dissipation), we estimated a short life for these disks with inner gaps 
(from few times $\sim$ 10$^5$ yr to less than 1 Myr; Paper III). 
Since here we find that accretion is present in only about half of the
``transition objects'', the same argument predicts an even shorter timescale
for stopping accretion or bringing it down to undetectable levels, 
once a gap has been opened in the inner disk. On the other hand, the
timescales for the opening of the gap remain uncertain, given the
lack of detailed studies of ``transition objects'' in younger and
older clusters. At this point, it is not possible to determine whether
the opening of the gap occurs rapidly in a normal CTTS-like disk, or
whether it is a longer process starting with the decrease in near-IR
emission with age. In Paper III we observed that more than 95\% of the
disks are below the median SED in Taurus, and that this difference
is more striking at shorter wavelengths. Tr 37 presents as well
some disks showing a remarkable ``kink'' in their SEDs, or an abbrupt
change in the SED slope occurring at about 8$\mu$m (see the cases of 73-472 and
11-2031 among others) that could be suggestive of a smooth evolution
into the ``transition object'' class. Nevertheless, detailed studies
of other populations are required to prove the timescales
of gap opening.

Even though the fraction of ``transition objects'' is small, it is too
large to be explained by the presence of a close stellar (or substellar) companion (which
would be otherwise a plausible mechanism to explain the inner
gaps without requiring disk evolution), taking into account the low close-in (few AU) binary fractions
for low-mass stars (Mathieu 1994; Lada 2006). Note that none of the ``transition
objects'' is found to be neither SB2 nor SB1. 
Photoevaporation of the inner disk by the central star
would be another method to produce inner gaps without involving
dust coagulation and/or planet formation \citep{clarke01}.
Photoevaporation could remove the gaseous and dusty component
of the inner disk in timescales comparable to the disk dissipation,
producing a very brief ($\sim$ 10$^5$ yr) phase during which the
inner disk is not present, before the outer disk is dissipated. 
The size of the resulting inner gap depends on the UV flux emitted by the star, 
so larger gaps are expected for earlier-type stars \citep{alexander06a,alexander06b}.
Further observations, oriented at the detection of photoevaporation and photodissociation of
gas in the outer (still optically thick) disk would be
required in order to determine if photoevaporation may be involved
in the disk dissipation. 

The presence of gas accretion in half of the ``transition objects''
may suggest grain growth and/or planet formation rather than photoevaporation
as the main cause for the near-IR gaps, as significant amounts of
gas need to remain in the disk and/or to flow through it. 
Grain growth to large sizes would result in a decrease of the
opacity at near- and mid-IR wavelengths, as we observe in about 90\% of the
disks in Tr 37 (Paper III), and grain growth to large (m or km) sizes would probably not affect
the gas flow in the disk. The formation of one or more planets would create a gap as well, and
depending on the planet and disk masses, slow accretion could occur for some time through the gap,
being able to reproduce both accreting and non-accreting ``transitional disks'' 
(Lin \& Papaloizou 1986; Bryden et al. 2000; D'Angelo et al. 2003; Quillen et al. 2004). 

Considering the limited spectral type range covered in this study, we do not
see any correlation between the spectral type and the presence and/or 
characteristics of the gaps. The ``transition objects'' we find in Tr 37
have spectral types ranging from K4 to M2, which is the
range containing most (96\%) of the studied objects. Despite the low
number of ``transition objects'' observed, we do not observe any tendency of ``transition
objects'' to have spectral type M, as pointed out by
McCabe et al. (2006). Note that the McCabe sample contains more
late-M stars than ours (which goes down to M2 stars), but
nearly 80\% of the objects observed by them are comparable to ours in
spectral type. Another point to be considered is that the sample
in McCabe et al. (2006) is younger ($\sim$ 1 Myr) than ours,
which is an important fact to be taken into account for timescale
estimates in the future.
The spectral type distribution of our ``transition
objects'' is consistent with random sampling in our population. 
We do not see either any correlation between
SEDs and spectral types, in the sense that most of these objects show photospheric colors
at wavelength shorter than $\lambda \sim$6 $\mu$m independently of 
their spectral types, and the excesses at or beyond 6$\mu$m do not show
any significant dependence with spectral type. Since our Spitzer study
is complete only down to 8 $\mu$m (the 24$\mu$m detection limit
only covers photosphere emission from B-type stars, and it
is about 1-2 orders of magnitude over the typical photosphere of a TTS; Paper III), 
the detection of gaps larger than $\sim$5 AU around low-mass stars is difficult or impossible
with this set of data, unless the outer disks are very luminous or we can infer the
presence of an outer undetected disk because of the presence of accretion. Therefore, our
study may be missing some non-accreting
``transitional objects'' with larger gaps. Note that the generalized sensitivity limitations for
detecting large gaps and/or ``debris disks'' around stars with
spectral types later than A may explain part of the bias towards
finding larger gaps around more massive stars. Here, the only case of a low-mass star with
a confirmed large gap is 13-819, which has no excess even at 8$\mu$m. It is 
confirmed to have some outer disk due to its accretion, detectable via both 
H$\alpha$ emission and U band excess. Given its spectral type (K5.5), similar
to other ``transition objects'' (like 13-566, 21402192+5730054), we do not find any correlation
between spectral type and size of the gap. 
Although the limitation both in number of objects and in the reduced spectral type range
covered by these observations must be taken into account, these data suggest
that photoevaporation alone is probably not able to reproduce the
set of observed ``transition objects'', and additional mechanisms
as grain growth to large (planetesimal) sizes and/or planet formation must be invoked.

Among the potential ``transition objects'' with larger gaps (even though they must be regarded with care due to the lack
of 24 $\mu$m data to confirm the presence of a disk), we would like to mention
the case of several stars showing a small excess at 8 $\mu$m only (typically, 7-10 sigma
over the photospheric emission, Paper III) and no 
broad H$\alpha$ line emission. The lack of 24 $\mu$m
counterparts suggest that emission at these wavelength is below our detection 
limit (keeping in mind that our detection limit is one to two orders of magnitude above
the photospheric emission of low-mass stars), as it happens in other stars with normal-looking
but relatively flat accretion disks (see 13-1891 and 12-1010 among others). 
The star 13-350 (M1) is one of these cases, having a very small excess at 5.8 $\mu$m,
in addition to a $\sim$10 sigma excess at 8$\mu$m, which makes it a good candidate
to be another non-accreting ``transition object''. Other stars with excesses only
at 8 $\mu$m are 14-222 (K7), 14-2148 (M1.5), 54-1613 (K5), and 92-1162 (M2). 
Both 14-222 and 54-1613 have uncertain very small U band excesses. As mentioned in Paper III, the
fact that this kind of objects is nearly absent from the older cluster NGC 7160,
suggests that at least part of them are true detections of ``transition objects''
and not photometry errors. Further followup of these objects
at longer wavelengths and with improved sensitivity would be recommendable, in
order to estimate the fraction (and therefore, timescales) 
of ``transition objects'' more accurately.

\subsection{Accretion rate evolution\label{accretion}}

Inspecting the H$\alpha$ profiles, we can complete our
knowledge about the accretion rates that we obtained from 
U band photometry (Gullbring et al. 1998). The accretion rate
calculations are described in detail in Paper II. There, we measured
the excess of U band luminosity over the photospheric level, related to
the accretion luminosity by the prescription in \citet{gullbring98}. 
U band excesses corresponding to accretion rates below 10$^{-9}$ M$_\odot$/yr 
are in general difficult to detect (based on our U band sensitivity for
K7-M2 stars; Paper II), but the presence of H$\alpha$ high velocity wings 
can be used as a direct proof of accretion or no accretion
even in those cases.  For objects with velocity wings broader than 200 km/s and
no detectable U excess, we assign 10$^{-9}$ M$_\odot$/yr as an upper limit
to the accretion rate. We define as 
non-accretors all these objects showing no U excess and/or
no U detection (our U photometry detects photospheric
U band emission only down to a K6-K7 star, see Paper II)
and no velocity wings broader than $\sim$ 200 km/s (White \& Basri 2003; Natta et al. 2004). 
According to the models by Muzerolle et al. (2003), accretion rates under
10$^{-12}$M$_\odot$/yr do not result in significant line
broadening, so these objects would have accretion rates
below 10$^{-12}$M$_\odot$/yr, if any. In the cases where no excess
from a disk is detected at any wavelength, the
accretion rate is likely to be zero or negligible.
Whenever no U band photometry is available, we base the
accretion/non accretion criterion on the presence of velocity wings.
In case of any doubts about the profiles, we prefer to 
be conservative, assigning an upper limit to the
accretion rate of $\sim$10$^{-9}$ M$_\odot$/yr. 
Table \ref{summary-table} contains the accretion rates from Paper II, completed with
the new upper limits for the sure and probable members.

\begin{figure}
\plotone{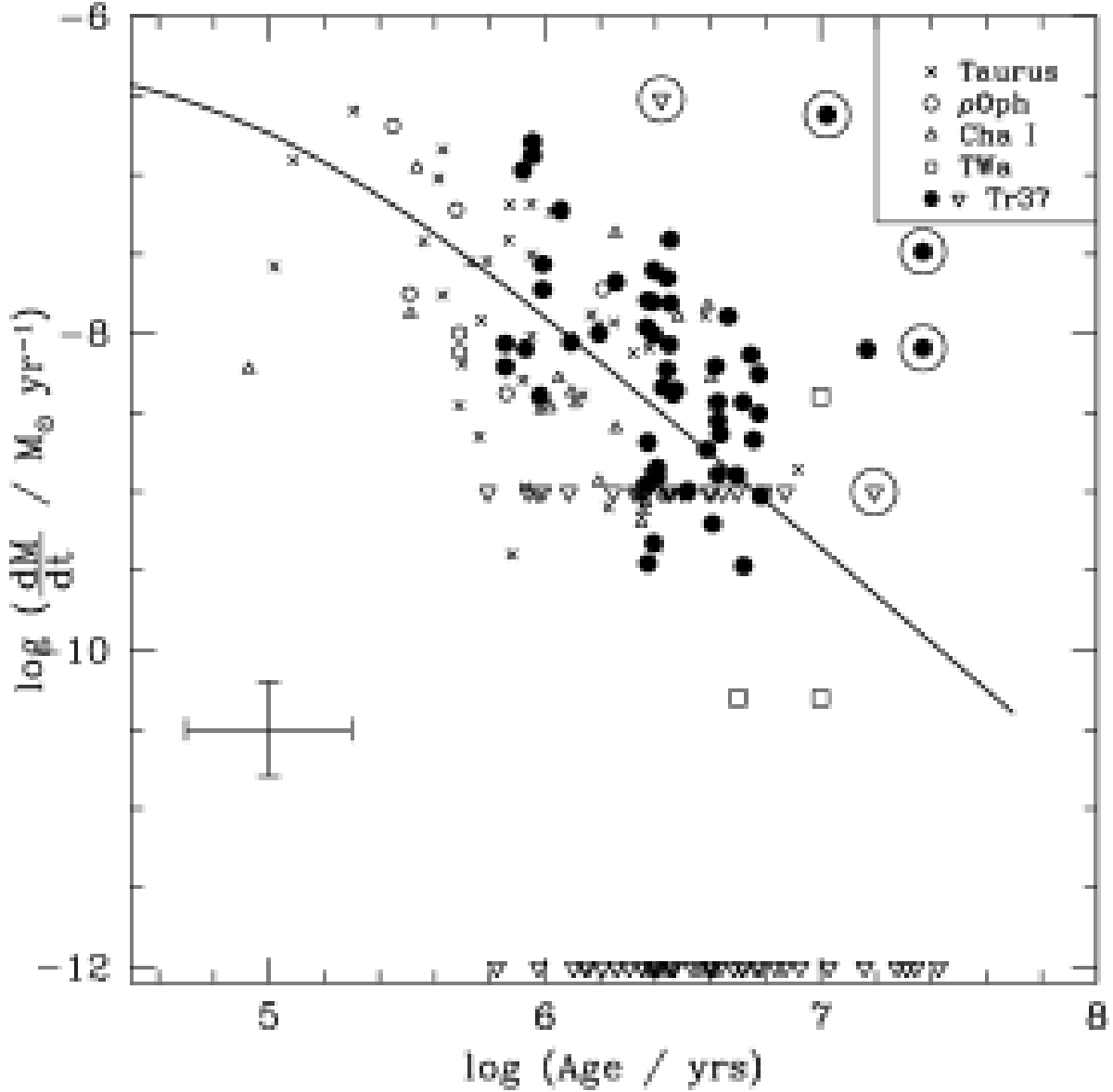}
\caption{Accretion rate vs. age in Tr 37. Filled circles represent
stars with accretion rates derived from U band (see Paper I, PaperII for
a discussion about the accretion rate estimates). Open triangles are
upper limits to the accretion based on the presence of broad H$\alpha$ emission
(10$^{-9}$M$_\odot$/yr) in stars for which we did not find any U excess (U band observations
were complete to approximately the U photospheric emission of a K6 star), and
the lack of broad components in stars with no U excess (10$^{-12}$M$_\odot$/yr). 
For comparison, we include data from other regions and 
the model for the evolution of a viscous disk (Hartmann et al. 1998; 
Muzerolle et al. 2000). The average accretion rate in Tr 37 (including
the upper limits, but excluding the G-type stars, marked here with large
open circles) is 9$\times 10^{-9}$ M$_\odot$/yr. The typical error bar
for \.{M} derived from U band and for the age is displayed.
\label{accrates} }
\end{figure}

Since the H$\alpha$ 
observations complete our information
about accretion processes in $\sim$ 80\% of all the
stars, we have revisited the accretion vs.
age diagram described in Paper II, including all the new
upper limits (Figure \ref{accrates}). Compared to the
evolution of a viscous disk (Hartmann et al. 1998; Muzerolle
et al. 2000), we find better evidence of the decrease of
accretion rate with time (using ages derived from the
dereddened V vs. V-I diagrams, which proved to be less affected by observational
errors than the HR diagram, and the Siess et al. 2000
isochrones; Paper II). The fraction of stars with accretion rates under
$10^{-9}$ M$_\odot$/yr is significantly larger than in younger clusters, and 
we can set an upper limit to
the median accretion rate of around 1$\times 10^{-9}$ M$_\odot$/yr
(including the $10^{-9}$ M$_\odot$/yr upper limits in the calculation).
The average rate is 9$\times$10$^{-9}$ M$_\odot$/yr (including
the $10^{-9}$ M$_\odot$/yr upper limits),
slightly lower than the 2$\times 10^{-8}$ M$_\odot$/yr
average obtained with the U band photometry alone, and
slightly lower than the Taurus average (10$^{-8}$ M$_\odot$/yr).
However, given that the standard deviation is around 
2.3$\times 10^{-8}$ M$_\odot$/yr because it depends mostly on 
a few very strong accretors, the median
rate is probably more significant (note that we have
excluded from these calculations the G-type stars, which
show systematically higher accretion rates).
The IR excess in Tr 37, which are systematically
lower than in Taurus (especially at shorter wavelengths, see
Paper III), the higher number of very slow accretors in Tr 37 compared to 
Taurus, and the fact that accretion rates are in general below the Taurus
average suggest that the evolution of the dust is somehow 
parallel to the gas and accretion evolution.
The fact that the ``transitional disks'' with inner IR gaps present
accretion rates never larger than $10^{-9}$ M$_\odot$/yr,
and consistent with \.{M}$<$10$^{-9}$ M$_\odot$/yr or even zero (Class III-like) accretion 
in about half of the cases, seems to indicate a correlation 
between dust and gas evolution as well.

\begin{figure}
\plotone{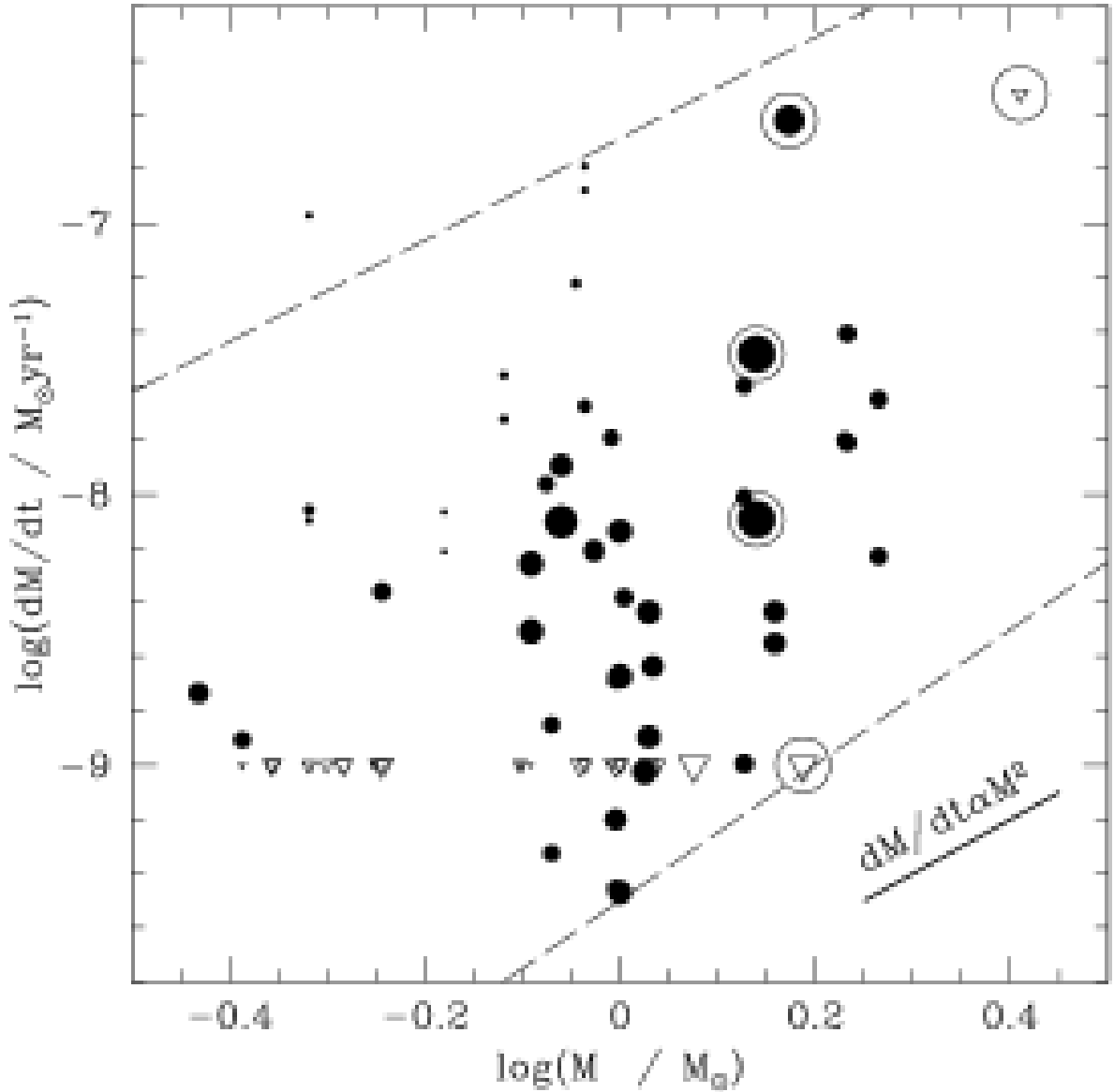}
\caption{Accretion rate vs. mass in Tr 37. The size of the dots and
open triangles (upper limits) is proportional to the age. Very small dots
represent the globule population (aged $\sim$1 Myr). Ages for G-type
stars tend to look larger ($\sim$ 10 Myr, G stars are marked by large open circles) but are highly
uncertain. The data is consistent with the study of Calvet et al. (2004), 
whose data is comprised between the dashed lines, and with the \.{M} $\alpha$ M$^2$ trend
observed by Natta et al. (2004), although our sample contains stars 
spanning a smaller parameter area in both accretion rate and mass. 
This trend appears independently of age in both the
globule and the bulk (4 Myr old in average) population, even though accretion
rates for the younger stars tend to be higher, as expected from the 
viscous disk evolutionary models (Hartmann et al. 1998).
\label{macc-mass} }
\end{figure}

Following Muzerolle et al. (2003), Natta et al. (2004),
and Calvet et al. (2004), we search for a correlation
between mass and accretion rate. As we did in Paper II, we
derived masses directly from the extinction-corrected
V vs. V-I diagram (see Table \ref{summary-table} for the
individual values of age and mass). Even though most of our
stars are roughly the same mass, we can see a slight evidence of a
\.{M} $\alpha$ M$^a$ trend in Figure \ref{macc-mass}. Comparing to the
sample in Calvet et al.(2004) and Natta et al.(2004), we find that
our stars are consistent with their samples, although our
reduced differences in M and in accretion rates here does not
make a \.{M} $\alpha$ M$^2$ relation evident. If we take into account the
age differences (even though the ages for G-type stars are
highly uncertain), we find that the stars belonging to the
globule population (which are $\sim$1 Myr old instead of
4 Myr, Paper II) show a similar trend, but with systematically
higher mass accretion rates, as we would expect from the viscous
disk evolutionary models (Hartmann et al. 1998; Muzerolle et al. 2000;
Figure \ref{accrates}).

\subsection{Rotation rates of accreting and non-accreting stars \label{rotation}}

The high-resolution spectra allow us to face the question concerning
the way stars are released from their inner disks during
the processes of planet formation and disk evolution. This could have
important consequences on the evolution of the disk and the angular
momentum of the star. There is
still a large controversy about the way accreting stars are connected to their
disks, and about the combined effects of contraction, accretion
and winds on the angular momentum of the star \citep{bouvier93,edwards93,stassun99,rhode01,kuker03}.
One interpretation suggests that the presence of an accreting disk 
would lock the star, preventing it from speeding up as it contracts and accretes 
\citep{bouvier93,shu94,choi96,herbst02,hartmann02}. Other observations suggest that
disk accretion and stellar contraction may affect
the stellar rotation differently, resulting in rotational
periods that do not change with time \citep{stassun01}. 
Some observations could be also consistent with a picture where the 
star-disk connection may not have any appreciable effects on the
rotation of the star, so the observed
differences would be mostly related to initial conditions rather
than to age or evolutionary stage \citep{stassun01}. It has been also
suggested that the effects of disk locking may be time-dependent, so disk locking could 
start only some time after the formation of the Class II system, 
or conversely, it could disappear with age as the accretion rates
drop down while the disk evolves \citep{hartmann02}. 
Nevertheless, the time-dependence of disk locking is highly
uncertain, given the lack of rotational studies of large populations 
with different ages. At the present moment, there is no clear answer,
as different studies result in different conclusions, and it is not known yet 
whether some historical problems identifying
accretion processes and/or the presence of disks may be responsible
for the variety of results \citep{sic05onc}, or whether this is a physical difference
resulting from different ages, initial conditions, environment or evolutionary 
stages. 

\begin{figure}
\epsscale{.9}
\plotone{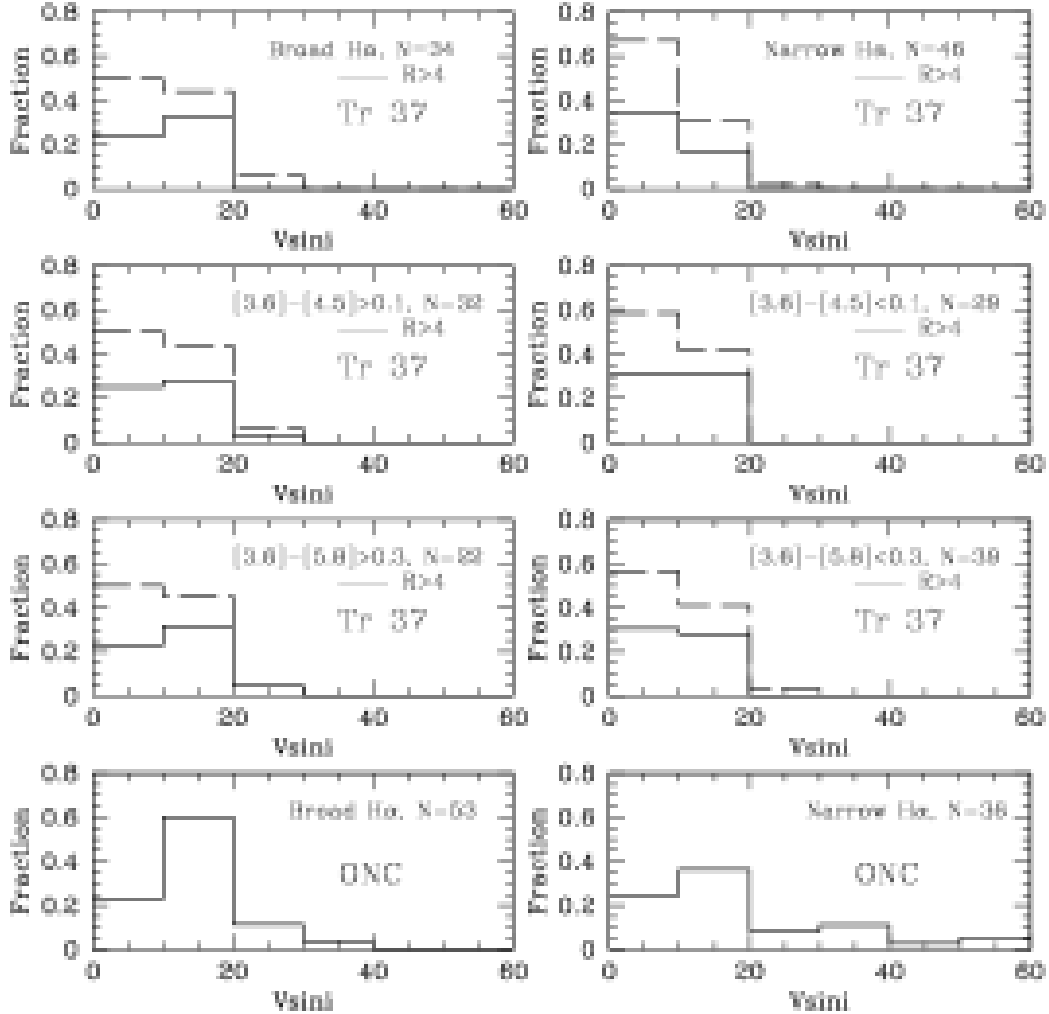}
\caption{Rotational velocities of stars with disks versus stars
 without disks. 
The fraction of stars in each velocity bin (Vsini in km/s) is
given for three different definitions of disk in Tr 37, displayed 
in the three upper rows of figures. Figures on the left correspond
to accreting and/or disked stars (typically, CTTS), figures on the right
correspond to non-accreting and/or diskless stars (typically, WTTS).
In the two upper pannels, we compare stars with broad and narrow H$\alpha$.
The second row of pannels defines the presence of a disk as
an excess in the IR such that the Spitzer color [3.6]-[4.5] is larger/
smaller than 0.1. The third row distinguishes stars with [3.6]-[5.8] colors
larger/smaller than 0.3 (see Paper III for an explanation of CTTS versus WTTS
typical colors). The fourth row shows the histograms for stars with broad versus
narrow H$\alpha$ emission in the ONC \citep{sic05onc}. We do not find
substantial difference in the rotation of CTTS vs. WTTS in Tr 37,
and the differences with the stars in the ONC may result from the differences
in the ages, cluster morphology, and also because of the effect of poorer
S/N (see text).
 \label{histo} }
\end{figure}

Here, we investigate the possible correlation between stellar
rotation (V$sin i$) and the presence of accretion and/or accretion disks. 
Tr 37 is a specially interesting cluster to study the effect
on rotation of disk locking because it is older than
previously studied young clusters, and younger than mid-aged
stars, which are no longer accreting.
We use the broad H$\alpha$ emission and the IR excesses (based on the [3.6]-[4.5]
and [3.6]-[5.8] IRAC colors, see Paper III)
to define the presence of ongoing accretion and of an inner
disk , respectively (for more
detail in the correlation of accretion and IR excesses, see Section \ref{HavsIR}). We have
constructed histograms for the rotational velocities V$sini$ for the stars with 
and without signatures of accretion and inner disk, separating them in 10 km/s bins
(in order to minimize the errors derived from the fiber tilts, 
Section \ref{observations}). Considering only the
stars with high signal-to-noise and good correlations, as well as good
IRAC/Spitzer photometry, our sample is reduced to
34 stars with broad H$\alpha$ vs. 46 stars with narrow H$\alpha$,
32 stars with [3.6]-[4.5] excess vs. 29 stars without this excess,
and 22 stars with [3.6]-[5.8] excess vs. 39 stars without [3.6]-[5.8] 
excess, which may mean low-number statistics. The different
histograms are shown in Figure \ref{histo}. From the comparison, we do not find
any significant differences in the rotation of CTTS and WTTS stars according
to any of these three definitions. Given the good correlation
of IR excess and broad H$\alpha$ profiles (Section \ref{HavsIR}), 
except for the few accreting ``transition objects'', all
the histograms are very similar.  
Even though these results are based on low-number statistics, 
the lack of significant differences could suggest 
that either disk locking is not an
important mechanism regulating the rotation of the stars, or it is only
important at earlier stages, when accretion
rates are higher and dust settling/grain growth is not so widely present.

Comparing with our previous study of the Orion Nebula Cluster
(Sicilia-Aguilar et al. 2005) we find relevant differences in the rotation of
WTTS (see Figure \ref{histo}) in the two regions. The fraction of WTTS 
with high rotational velocities is significantly higher in the ONC than in Tr 37, and the
distribution of rotational velocities of CTTS and WTTS in the ONC are
clearly different \citep{sic05onc}. The ONC and Tr 37 
differ in age ($\sim$ 1 Myr versus $\sim$ 4 Myr 
respectively), disk fraction ($\sim$ 80\% versus $\sim$ 45\%) and
environment (the ONC being much denser, with more massive stars, more embedded in the 
original nebula, and more populous than Tr 37),
so it is not clear that their star formation and disk evolution histories
can be simply compared. Even though the differences in rotation
between Tr 37 and the ONC may
suggest some degree of rotational evolution during the first million years
of the life of a star, the observed variations may just show a dependence
on initial and/or environmental conditions, as suggested by Stassun et al. (2001).

\section{Conclusions \label{conclu}}

We presented a study of the accretion processes occurring in 
the 4 Myr-old cluster Tr 37, using the high-resolution
spectrograph Hectochelle in the MMT telescope, to observe the
H$\alpha$ order. Observing a total of 460
stars, we confirmed the membership and accretion properties
of 144 previously known members, and we found a total of 26
new members and probable members, according to their H$\alpha$ emission and to
their radial velocities. We use H$\alpha$ as a
powerful sign of accretion (characterized by asymmetric profiles
and broad velocity wings), enabling us to detect stars
with accretion rates well below 10$^{-9}$ M$_\odot$/yr and up to
10$^{-12}$ M$_\odot$/yr, which are below the detection limits of U band
surveys. This way, we are able to set limits to accretion
to a large number of stars in the cluster, finding that the
average accretion rate is lower than in younger regions (i.e.
Taurus), being of the order of 9$\times$10$^{-9}$ M$_\odot$/yr.
Given the similarities between the decrease in IR excesses
observed in Tr 37 with respect to Taurus, and the decrease
in accretion rates, gas evolution seem to occur
somehow parallel to the evolution of the dust grains and
the structure of the disk. 

We find that ``transition objects'', or stars with inner
gaps in their disks (seen as objects with no IR excess at 
wavelength shorter than $\sim$6 $\mu$m) show either no accretion
($<$10$^{-12}$ M$_\odot$/yr)
or are consistent with extremely low accretion rates, mostly
undetectable from U band photometry (which sets an upper limit
of 10$^{-9}$ M$_\odot$/yr).  The ``transition objects'' represent 
about 10\% of the total number of disks observed in Tr 37, and 
about half of them do not show any evidence of accretion. The
number of ``transition objects'' observed here is large than in other
regions, although this study is more complete than others since it
includes observations up to 24 $\mu$m and high-resolution spectra, so
we are able to identify some objects looking as intermediate stages
between class II and class III as ``transition objects''.
Additionally,
accretion rates seem much lower (or undetectable) in the potential disks 
with larger gaps, even though we find 1 case of accretion in a disk with a
large inner gap (presumably $>$5-10 AU).
The decrease and termination of the accretion 
processes as the inner disk clears up/agglomerates
suggest that the evolution of dust and gas occurs in a parallel way,
and that mass accretion may cease shortly after the formation of these
inner disk gaps. Therefore, these objects are a key to the intermediate
stage between CTTS and WTTS, in which accretion processes are
terminating and important planet formation may be taking place.

Even though the sample of ``transition objects'' is limited in number
and in spectral type coverage,
the fact that half of the ``transition objects'' are accreting
may suggest grain growth and planetesimal (or even planet) formation
as the cause for inner gaps in disks, at least in part of the objects.
We do not observe any differences in the size of the gap according 
to the spectral type, and the presence of 1 confirmed and several probable
disks with large gaps seem to suggest that stars with the same
spectral type can have inner holes of different
sizes. These observations would suggest that grain growth/settling is responsible
for the opening of inner gaps, rather than photoevaporation alone. 
Observations at longer wavelengths would be advisable in order to
confirm the larger gaps and the presence of rims and walls, and UV line 
observations could be used to determine the presence
of photoevaporation in the rim of the outer disk of ``transition objects''.

Finally, we obtain the rotational velocities for about 75\% of the stars, 
finding no significant differences
between the rotation of accreting and non-accreting stars.
Therefore, disk locking is either not relevant at the
age of 4 Myr, or it is dependent of the environment, 
or the differences observed in other regions
account rather for different initial conditions than for 
the evolutionary stage.

We want to acknowledge J. Muzerolle, S. Mohanty and B. Mer\'{\i}n
for the interesting comments and discussion. We also want to thank
the anonymous referee for the detailed review of our paper and useful comments.
This publication makes use of data products from the Two Micron All Sky Survey, 
which is a joint project of the University of Massachusetts and the Infrared 
Processing and Analysis Center/California Institute of Technology, funded by the 
National Aeronautics and Space Administration and the National Science Foundation. 
This work made use of the VizieR Astronomical Database.

\appendix

\section{APPENDIX: The star 13-277}

A special case that deserves attention is the star 13-277. 
It is classified as a late G/early K star (the spectral type is uncertain
due to high veiling), and it has an extremely bright disk, 
plus extremely high luminosity for a late-type star (L$\sim$26 L$_\odot$).
Its accretion rate, although uncertain, is much higher than the average
accretion rate in Tr 37, being approximately $\sim$3 $\times 10^{-7}$ 
M$_\odot$/yr according to its 10\% H$\alpha$ EW (see Natta et al. 2004), and
maybe up to 10$^{-5}$ M$_\odot$/yr (depending on the actual spectral type)
calculated from the U band excess. It is by far the fastest rotator in the
cluster (V$sini \sim$52 km/s, the next fastest rotator has V$sini\sim$28 km/s). 
Its radial velocity is slightly off-cluster (with a deviation slightly higher
than 2 $\sigma$), but this is most likely due to the difficulties obtaining
accurate radial velocities when the lines are very broadened by rotation.

This star could be the object named G$\mu$ Cep and noted as a long-period
variable by Morgenroth (1939). There is an offset of about 1 arcmin in declination
between 13-277 and G$\mu$ Cep, but the position cited by Morgenroth (1939)
does not correspond to any other object. In that case, the photometry from
the Sonneberg Observatory reveals variations up to 2 magnitudes, although it
does not cite any kind of periodicity nor timescale for the star. At the present
time, 13-277 would be about its maximum magnitude cited in Morgenroth (1939).
Given that high accretion, extreme brightness, and fast rotation
are some of the characteristics of FU Ori and for EX Ori objects (Hartmann \& Kenyon 1996; Herbig et al. 2003;
Vittone \& Errico 2005), and that its disk is comparable to the 
brightest CTTS disk known (GW Ori; Mathieu et al. 1995), we
are planning an optical and radio followup of this object in order to determine
its true nature.

\clearpage

\clearpage

\pagestyle{empty}

\begin{center}
\begin{deluxetable}{llccccccccl}
\tabletypesize{\footnotesize}
\rotate
\tablenum{1}
\tablecolumns{11} 
\tablewidth{0pc} 
\tablecaption{Spectral Information \label{spec-dat}} 
\tablehead{
 \colhead{ID1} &\colhead{2MASS ID} & \colhead{Sp.Type} &\colhead{EW(H$\alpha$ \AA)} & \colhead{EW(Li 6707 \AA)} & \colhead{H$\alpha$ Type}
&\colhead{$H\alpha$ 10\% V(km/s)} &\colhead{R} &\colhead{cz (km/s)} &\colhead{Vsini (km/s)} &\colhead{Comments}}
\startdata
71-1309 & 21340974+5729550 & --- & --- & --- & w &    ---      & 2.0: & -22.7$\pm$2.1: & 8.4: & PN(r)* \\ 
74-48 & 21344730+5731148 & --- & --- & --- & w &    ---   & 4.2 & -8.5$\pm$1.1 & 7.6 & P(r)* \\ 
73-758 & 21350835+5736028 & K6.5 & (-9) & 0.4 & w: &   ---   & 2.6 & 1.7$\pm$1.6 & 7.7 & Y(e),N(r),TOn: \\ 
72-1427 & 21351627+5728222 & M1 & -16(-77) & --- & c &   330      & 1.5: & 122.0$\pm$1.7: & 5.6: & Y(e),N(r) \\ 
81-541 & 21351745+5748223 & K5.5 & -31(-60) & 0.4 & c &   380      & --- & --- & --- & Y(e) \\
64-156 & 21351804+5709441 & --- & --- & --- & w &    ---   & 3.7 & -8.1$\pm$1.2 & 7.4 & P(r)* \\ 
73-472 & 21351861+5734092 & K5 & -23(-10) & 0.7 & c & 300        & 3.4 & -17.1$\pm$1.7 & 9.6 & Y(e),Y(r) \\ 
73-674 & 21352076+5735288 & --- & --- & --- & w &   ---    & 9.3 & -24.5$\pm$0.5 & 7.2 & PN(r)* \\ 
73-1059 & 21352386+5738145 & --- & --- & --- & w &   ---   & 5.3 & -12.8$\pm$1 & 8.6 & Y(r)* \\ 
73-311 & 21352451+5733011 & M1.5 & (-29) & --- & c &    ---      & --- & --- & --- & Y(e) \\
73-71 & 21353021+5731164 & K6 & -10(-12) & 0.4 & c &   380      & 3.7 & -11.5$\pm$3.6 & 24.4 & Y(e),Y(r) \\ 
72-875 & 21354975+5724041 & M0.5 & (-21) & 0.1: & w(c) &    ---   & 1.6 & 17.3$\pm$2 & 7 & Y(e),N(r),SB1: \\ 
61-608 & 21355070+5703570 & --- & --- & --- & w &   ---    & 4.4 & -15.6$\pm$1.2 & 8.2 & Y(r)* \\ 
64-376 & 21355082+5712071 & --- & --- & --- & w &   ---   & 3 & -22.2$\pm$1.5 & 7.8 & PN(r)* \\ 
73-194 & 21355223+5732145 & K6.5 & (-4) & --- & w &   ---     & 3.4 & -14.4$\pm$1.2 & 7.3 & P(e),Y(r),TOn \\ 
61-893 & 21360090+5707129 & --- & --- & --- & w &    ---   & 2.3: & -9.8$\pm$2.2: & 9.9: & P(r)* \\ 
73-537 & 21360723+5734324 & G1.5 & (-5) & 0.3 & w(c) &   ---    & 3.1 & -20.5$\pm$2.2 & 11.7 & Y(e),P(r) \\ 
84-23 & 21361281+5753004 & --- & --- & --- & w &  ---    & 3.8 & -18.3$\pm$1.5 & 9.6 & Y(r)* \\ 
--- & 21362368+5732452 & --- & --- & --- & w &    ---  & 1.6: & -11.7$\pm$2.3: & 7.7: & Y(r)* \\ 
--- & 21362507+5727502 & M0 & -78:(-86) & --- & c &    ---      & 2.4 & -68.9$\pm$1.7 & 7.6 & Y(e),N(r),SB1 \\ 
61-413 & 21362615+5701293 & --- & --- & --- & w &     ---     & 3.2 & -25.6$\pm$0.9 & 5.2 & PN(r)* \\ 
14-306 & 21362676+5732374 & K6.5 & (-8) & --- & w &    ---      & --- & --- & --- & P(e) \\
--- & 21364596+5729339 & ClassI & -47 & --- & w:(c) &   ---       & 1.4: & -11.8$\pm$1.8: & 5.7: & Y(e),Y(r) \\ 
--- & 21364762+5729540 & K6 & (-76) & 0.5 & w(c) &     ---     & --- & --- & --- & P(e) \\
14-141 & 21364941+5731220 & K6 & -15(-5) & 0.6 & c &   570      & 2.3: & -9.2$\pm$3.6: & 27.0: & Y(e),P(r) \\ 
14-1229 & 21365579+5736533 & K6 & (-5) & 0.4 & w &    ---      & --- & --- & --- & Y(e) \\
11-2146 & 21365767+5727331 & K6 & -28(-33) & 0.5 & c &  470       & 7.6 & -15.4$\pm$1.1 & 12.6 & Y(e),Y(r) \\ 
11-1209 & 21365850+5723257 & K6 & -6(-4) & 0.7 & c(w) &   350      & 10 & -8.4$\pm$0.8 & 12.2 & Y(e),P(r) \\ 
--- & 21365947+5731349 & M0 & -30(-109) & --- & c &   510      & --- & --- & --- & Y(e) \\  
11-1659 & 21370088+5725224 & K5 & (-2) & 0.5 & w &    ---      & 7.7 & -14.6$\pm$0.8 & 9.1 & Y(e),Y(r) \\ 
11-1499 & 21370140+5724458 & M1.5 & (-7) & 0.3: & w &   ---       & --- & --- & --- & Y(e) \\
11-2322 & 21370191+5728222 & M1 & -18(-23) & 0.4 & c &  420       & 2.7: & -9.9$\pm$2.0: & 9.5: & Y(e),P(r) \\ 
11-1871 & 21370254+5726144 & M2 & (-13) & 1 & w(c) &   ---       & --- & --- & --- & Y(e) \\
14-222 & 21370607+5732015 & K7 & (-5) & 0.6 & w &     ---     & --- & --- & --- & Y(e),TOn: \\
14-287 & 21370649+5732316 & M0 & -18(-35) & 0.5 & c &   470      & 2.9 & -15.7$\pm$2.1 & 11 & Y(e),Y(r) \\ 
11-2037 & 21370703+5727007 & K4.5 & -43(-50) & 0.5 & c &  420       & 9.8 & -19.9$\pm$0.6 & 8.6 & Y(e),P(r) \\ 
11-1067 & 21370843+5722484 & M0.5 & (-7) & --- & w &   ---       & --- & --- & --- & Y(e) \\
14-11 & 21371031+5730189 & M1.5 & (-5) & 0.3 & w &    ---      & 2 & -15.0$\pm$3.7 & 15.1 & Y(e),Y(r),TOn \\ 
14-125 & 21371054+5731124 & K5 & -14(-13) & 0.6 & c &  320      & 2.8 & -16.7$\pm$1.1 & 5.6 & Y(e),Y(r) \\ 
11-1513 & 21371183+5724486 & K7.5 & (-5) & 0.5 & w &   ---       & --- & --- & --- & Y(e) \\
11-2131 & 21371215+5727262 & K6.5 & (-10) & 0.5 & c &   ---    & --- & --- & --- & Y(e) \\
11-2487 & 21371498+5729123 & K7 & (-4) & 0.4 & w &   ---       & --- & --- & --- & Y(e) \\
11-2031 & 21371591+5726591 & K2 & -5(-5) & 0.4 & c &  440       & 6.8 & -17.2$\pm$1.2 & 12.6 & Y(e),Y(r) \\ 
14-103 & 21371976+5731043 & K7 & (-2) & 0.3 & w &   ---       & --- & --- & --- & Y(e) \\
14-197 & 21372368+5731538 & K5.5 & (-2) & 0.5 & w &    ---      & 2.1: & -25.0$\pm$3.9: & 20.4: & Y(e),PN(r),SB1: \\ 
--- & 21372410+5724115 & --- & -60 & --- & c &   430      & --- & --- & --- & Y(e)* \\
14-160 & 21372732+5731295 & K5 & (-22) & 0.6 & c:(c) &    ---    & 4.3 & -16.7$\pm$2.9 & 22.5 & Y(e),Y(r) \\ 
11-581 & 21372828+5720326 & G & (-9) & --- & w &    ---      & 6.7 & -68.2$\pm$0.7 & 6.7 & P(e),N(r),rej \\ 
14-1017 & 21372894+5736042 & M0 & (-55) & 0.5 & c &    ---      & --- & --- & --- & Y(e) \\
14-335 & 21372915+5732534 & K6.5 & (-20) & 0.6 & w:(c) &    ---   & --- & --- & --- & Y(e) \\
11-1864 & 21373420+5726154 & G-K & (-5) & 0.3 & w(c) &    ---   & 3.1 & -42.8$\pm$1.7 & 9.5 & Y(e),N(r),SB1 \\ 
83-343 & 21373696+5755149 & M0.5 & (-3) & 0.5 & c:(w) &    ---   & 3.5 & -21.3$\pm$1.5 & 8.8 & Y(e),P(r) \\ 
14-183 & 21373849+5731408 & K7.0(K5) & -14(-65) & 1.2: & c &   490      & 2.2 & -22.4$\pm$3 & 12.8 & Y(e),PN(r),SB1: \\ 
14-995 & 21373987+5736029 & --- & --- & --- & w &    ---   & 1.9: & -7.5$\pm$1.6: & 6.0: & PN(r)* \\ 
14-2148 & 21374184+5740400 & M1.5 & (-2) & 0.4 & w &   ---    & 1.8: & -21.3$\pm$1.3: & 4.8: & Y(e),P(r),TOn: \\ 
--- & 21374275+5733250 & F9 & -32(-54) & 0.1: & c &    380     & --- & --- & --- & Y(e) \\
11-1384 & 21374486+5724135 & K6.5 & (-5) & 0.6 & w &     ---     & --- & --- & --- & Y(e) \\
11-383 & 21374514+5719423 & K5 & -37(-23) & 0.3 & c &  270       & --- & --- & --- & Y(e) \\
--- & 21374893+5723209 & --- & -42 & --- & c &   540      & --- & --- & --- & Y(e)* \\
13-924 & 21375018+5733404 & K5 & (-4) & 0.6 & w &   ---       & 3.4 & -20.3$\pm$1.5 & 8.8 & Y(e),P(r) \\ 
12-1984 & 21375022+5725487 & K6 & (-5) & 0.7 & w &   ---       & --- & --- & --- & Y(e) \\
12-2519 & 21375107+5727502 & K5.5 & (-8) & 0.5 & w(c) &    ---      & 2.6 & -18.5$\pm$2.2 & 10.5 & Y(e),Y(r) \\ 
12-1968 & 21375487+5726424 & K6 & -8(-11) & 0.4: & c &  390       & 6.9 & -16.9$\pm$1 & 10.9 & Y(e),Y(r),SB2: \\ 
12-1422 & 21375756+5724197 & M0 & (-17) & 0.7 & w(c) &    ---      & --- & --- & --- & Y(e) \\
12-1091 & 21375762+5722476 & G2.5 & -2(-17) & 0.5 & c &    ---      & 6.1 & -15.8$\pm$0.9 & 8.8 & Y(e),Y(r) \\ 
13-269 & 21375812+5731199 & K6.5 & (-7) & 0.4 & w &  ---     & --- & --- & --- & Y(e) \\
12-583 & 21375827+5720354 & M0 & (-7) & --- & w &    ---   & --- & --- & --- & Y(e) \\
12-94 & 21375841+5718046 & K4 & (-4) & --- & w &    ---   & 7.7 & -117.9$\pm$0.6 & 6.4 & P(e),N(r),rej \\ 
12-94 & 21375841+5718046 & K4 & (-4) & --- & w &    ---   & --- & --- & --- & P(e) \\
13-1143 & 21375852+5735479 & --- & --- & --- & w &   ---    & 5.7 & -11.2$\pm$0.8 & 7.2 & P(r)* \\ 
13-1238 & 21375926+5736162 & M1 & -31(-64) & 0.7 & c &   550      & --- & --- & --- & Y(e) \\
12-2373 & 21380058+5728253 & M1 & (-6) & 0.4 & w &   ---   & 2.6 & -16.7$\pm$2.1 & 10 &Y(e), Y(r) \\ 
82-272 & 21380350+5741349 & G9 & -15(-13) & 0.4 & c &         & 2.6 & -15.6$\pm$3.8 & 18.1 & Y(e)Y(r),SB2 \\ 
12-1081 & 21380593+5722438 & M0.5 & (-4) & 0.7 & w &   ---   & --- & --- & --- & Y(e) \\
13-1161 & 21380772+5735532 & M0 & (-1) & 0.5 & w &   ---    & 2.2: & -21.8$\pm$3.0: & 12.8: & Y(e),P(r) \\ 
12-1613 & 21380848+5725118 & M1 & (-13) & --- & w(c:) &  --- & --- & --- & --- & P(e) \\
13-1426 & 21380856+5737076 & M0 & -40(-109) & --- & c &   460      & --- & --- & --- & Y(e) \\
--- & 21380924+5720198 & --- & -17 & --- & c &   490      & 1.6: & -22.0$\pm$2.4: & 8.2: & Y(e),P(r)* \\ 
13-669 & 21380928+5733262 & K1 & -18(-22) & 0.6 & c &   500      & 9.3 & -21.3$\pm$1 & 14.2 & Y(e),P(r) \\ 
--- & 21380979+5729428 & --- & --- & --- & w: &    ---   & 2.9 & -17.3$\pm$1.7 & 9 & Y(r)* \\ 
13-838 & 21381120+5734181 & --- & --- & --- & w &   ---    & 2.1 & -6.3$\pm$1.6 & 6.6 & PN(r)* \\ 
13-350 & 21381384+5731414 & M1 & (-9) & --- & w &   ---    & 2.9 & -12.6$\pm$1.2 & 6.4 & P(e),Y(r), conf,TOn:\\ 
13-350 & 21381384+5731414 & M1 & (-9) & --- & w &   ---   & 1.9: & -16.9$\pm$1.8: & 7.1: & P(e),Y(r), conf,TOn: \\
12-1017 & 21381509+5721554 & K5.5 & (-4) & 0.6 & w &   ---    & 9.8 & -15.1$\pm$0.6 & 8.8 & Y(e),Y(r) \\ 
54-1781 & 21381612+5719357 & M1 & (-13) & 0.3: & c:(c) &    ---   & 2.1 & -15.6$\pm$1.7 & 7.1 & Y(e),Y(r) \\ 
54-1781 & 21381612+5719357 & M1 & (-13) & 0.3: & c:(c) &    ---  & 1.7: & -13.4$\pm$2.2: & 8.1: & Y(e),Y(r) \\ 
13-1877 & 21381703+5739265 & K7 & -33(-68) & 0.4 & c &  430       & --- & --- & --- & Y(e) \\
13-277 & 21381731+5731220 & G1 & -14(-10) & --- & c &   660      & 4.5 & -7.5$\pm$6.7 & 51.6 & Y(e),PN(r),SB1: \\ 
13-2236 & 21381749+5741019 & K6.5 & (-1) & 0.6 & w &   ---   & 5.4 & -13.8$\pm$1.6 & 15.2 & Y(e),Y(r) \\ 
12-1009 & 21381750+5722308 & K5.5 & (-4) & 0.6 & w &   ---   & 3.7 & -18.4$\pm$2 & 12.6 & Y(e),Y(r) \\ 
94-1119 & 21381862+5803283 & --- & --- & --- & w &    ---  & 11.2 & -19.0$\pm$0.5 & 7.3 & P(r)* \\ 
13-819 & 21382596+5734093 & K5.5 & -6(-10) & 0.5 & c &   ---     & --- & --- & --- & Y(e),TOa \\
94-1050 & 21382668+5802377 & --- & --- & --- & w &   ---   & 2 & -13.1$\pm$1.9 & 7.8 & Y(r)* \\ 
12-1955 & 21382692+5726385 & K6.5 & (-2) & 0.4 & w &   ---    & 7.5 & -19.1$\pm$0.8 & 9 & Y(e),P(r) \\ 
13-236 & 21382742+5731081 & K2 & -56(-47) & --- & c &   500      & 9.3 & -18.2$\pm$0.7 & 9 & Y(e),Y(r) \\ 
12-2113 & 21382743+5727207 & K6 & -15(-28) & 0.4 & c &   430      & 3.7 & -26.1$\pm$1.8 & 11.7 & Y(e),N(r),SB1 \\ 
13-157 & 21382804+5730464 & K5.5 & -20(-14) & 0.5 & c &   560      & 3.8 & -15.6$\pm$1.3 & 8.7 & Y(e),Y(r) \\ 
13-232 & 21382834+5731072 & M0 & (-3) & 0.3: & w &   ---    & --- & --- & --- & Y(e) \\
--- & 21383216+5726359 & M0 & (-129) & 0.3: & c &   ---   & --- & --- & --- & P(e) \\
13-52 & 21383255+5730161 & K7 & (-1) & 0.7 & w &    ---   & 3.2 & -19.1$\pm$1.5 & 8.1 & Y(e),P(r),TOn \\ 
12-1825 & 21383382+5726053 & --- & --- & --- & w &    ---    & 2.4 & -15.8$\pm$2 & 8.8 & Y(r)* \\ 
91-155 & 21383470+5741274 & M2.5 & (-8) & 0.4 & w &   ---   & 1.5 & -9.3$\pm$1.9 & 5.9 & Y(e),P(r) \\ 
13-566 & 21383481+5732500 & K5.5 & (-5) & 0.4 & w &   ---  & --- & --- & --- & Y(e),TOn \\
13-1891 & 21384001+5739303 & M0 & (-11) & --- & c &   ---    & 3.3 & -13.8$\pm$1.8 & 10.1 & P(e),Y(r),conf \\ 
13-1709 & 21384038+5738374 & K5.5 & (-3) & 0.4 & w &  ---    & 3.3 & -15.6$\pm$1.8 & 10.4 & Y(e),Y(r) \\ 
54-1613 & 21384332+5718359 & K5 & (-1) & 0.5 & w &   ---    & 12.3 & -16.5$\pm$0.5 & 9.5 & Y(e),Y(r),TOn: \\ 
--- & 21384350+5727270 & M2 & (-23) & 0.4 & w(c) &   ---   & --- & --- & --- & Y(e),TOn \\
54-1547 & 21384446+5718091 & K5.5 & -33(-34) & 0.4 & c &  540       & 1.8 & -15.7$\pm$4.9 & 20 & Y(e),Y(r) \\ 
12-2363 & 21384544+5728230 & M0.5 & (-3) & 0.6 & w &   ---    & 2.7 & -7.4$\pm$4.7 & 27.1 & Y(e),PN(r),SB1:\\
12-595 & 21384622+5720380 & K7 & (-17) & 1.6 & w(c) &   ---   & --- & --- & --- & Y(e),TOn: \\
12-1423 & 21384707+5724207 & K7 & (-2) & 0.5 & w &    ---  & 4 & -15.6$\pm$1.1 & 7.5 & Y(e),Y(r) \\ 
12-1010 & 21385029+5722283 & M2 & -23(-20) & 0.2: & c &   270      & 2.5 & -16.5$\pm$1.8 & 8.2 & Y(e),Y(r) \\ 
12-2098 & 21385253+5727184 & M2.5 & (-7) & 0.7 & w &   ---   & --- & --- & --- & Y(e) \\
21-851 & 21385504+5720423 & --- & --- & --- & w &    ---  & 1.5 & -17.5$\pm$1.8 & 6 & Y(r)* \\ 
13-1087 & 21385542+5735299 & K4 & (-2) & 0.6 & w &   ---   & 12.7 & -15.0$\pm$0.6 & 11.2 & Y(e),Y(r) \\ 
91-506 & 21385807+5743343 & K6.5 & -31(-47) & 0.6 & c &  450       & 3.1 & -15.8$\pm$1.5 & 8.4 & Y(e),Y(r) \\ 
24-692 & 21390054+5734280 & M1 & (-2) & 0.5 & w &   ---   & 2.1 & -35.0$\pm$1.5 & 6 & Y(e),N(r),SB1 \\ 
12-1027 & 21390319+5722318 & M0 & (-11) & 0.8 & w(c) &    ---    & --- & --- & --- & Y(e) \\
24-77 & 21390346+5730527 & K6.5 & (-3) & 0.5 & w &  ---    & --- & --- & --- & Y(e) \\
24-108 & 21390390+5731037 & K5.5 & (-3) & 0.4 & w &   ---   & --- & --- & --- & Y(e) \\
12-1617 & 21390468+5725128 & M1 & -13(-30) & 0.6 & c &  ---    & 1.7: & -13.4$\pm$1.7: & 6.1: & Y(e),Y(r) \\ 
12-1650 & 21390471+5725215 & --- & --- & --- & w &   ---    & 6.7 & -25.2$\pm$0.8 & 7.8 & PN(r)* \\ 
13-1048 & 21391088+5735181 & M0 & -8(-7) & 0.4 & c(w) &   430      & 6.6 & -16.1$\pm$1.1 & 10.9 & Y(e),Y(r) \\ 
13-1250 & 21391213+5736164 & K4.5 & -4(-2) & 0.4 & c(w) &   540      & 9.7 & -15.7$\pm$0.6 & 8.3 & Y(e),Y(r),TOa \\ 
21-563 & 21391288+5721088 & M1 & (-2) & 0.3 & w(c) &   ---   & --- & --- & --- & Y(e) \\
12-942 & 21391424+5722129 & K7.5 & (-4) & 0.7 & w &    ---   & 4.1 & -4.0$\pm$2.7 & 19.3 & Y(e),N(r),SB1 \\ 
91-815 & 21391465+5745177 & M2 & (-4) & 0.2: & w &   ---   & 2 & -17.1$\pm$1.6 & 6.5 & Y(e),Y(r) \\ 
21-1590 & 21391554+5726440 & K7 & (-4) & 0.6 & w &    ---  & 4.3 & -17.3$\pm$1.5 & 10.6 & Y(e),Y(r) \\ 
21-1189 & 21391583+5724350 & --- & --- & --- & w &   ---   & 2.1: & -7.0$\pm$2.0: & 8.7: & PN(r)* \\ 
--- & 21392541+5733202 & --- & --- & --- & c &   200      & 8.1 & -14.4$\pm$0.7 & 8.8 & Y(r),TOa \\ 
--- & 21392570+5729455 & --- & --- & --- & w: &    ---   & 2.2: & -15.0$\pm$1.7: & 7.5: & Y(r)*,TOa \\ 
24-542 & 21392957+5733417 & K4 & (-4) & 1.3 & w &   ---   & --- & --- & --- & Y(e) \\
--- & 21393104+5747140 & --- & -16 & --- & c &   450      & 15.5 & -15.8$\pm$0.4 & 8.8 & Y(e),Y(r)* \\ 
24-515 & 21393407+5733316 & M0.5 & (-11) & --- & c &   ---    & 1.9: & -19.4$\pm$2.0: & 7.6: & Y(e),P(r),TOa \\ 
21-998 & 21393480+5723277 & K5.5 & (-16) & 0.3: & c &   ---    & 1.6 & -78.8$\pm$2.1 & 7.3 & Y(e),N(r),SB1: \\ 
21-33 & 21393561+5718220 & M0 & -51(-107) & --- & c &  450       & 1.8: & -13.2$\pm$2.4: & 8.9: & Y(e),Y(r) \\ 
24-170 & 21393612+5731289 & K7.5 & (-4) & 0.4 & w &    ---   & 1.5: & -14.4$\pm$2.2: & 7.0: & Y(e),Y(r) \\ 
53-1803 & 21393803+5719332 & K6.5 & (-1) & 0.4 & w &   ---    & 3.5 & -107.9$\pm$1.2 & 7.2 & Y(e),N(r),SB1 \\ 
24-48 & 21393805+5730439 & M0.5 & (-1) & 0.5 & w &   ---   & 6.2 & -15.1$\pm$0.8 & 8.1 & Y(e),Y(r) \\ 
92-1198 & 21394009+5746561 & --- & --- & --- & w &    ---   & 7.7 & -17.0$\pm$1.2 & 14.1 & Y(r)* \\ 
21-230 & 21394169+5719274 & M0.5 & (-3) & 0.5 & w &    ---   & 4.3 & -14.3$\pm$1.6 & 11.3 & Y(e),Y(r) \\ 
92-393 & 21394408+5742159 & M2 & -21(-34) & --- & c &   520      & 2.4 & -15.8$\pm$2.9 & 14.2 & Y(e),Y(r),TOa \\ 
24-820 & 21394746+5735059 & K6.5 & (-2) & --- & w &   ---    & 4.1 & -11.0$\pm$1 & 6.9 & P(e),P(r),conf \\ 
21-2251 & 21394754+5725210 & M2 & (-4) & 0.4: & w &   ---   & 1.8: & -2.6$\pm$4.4: & 16.7: & Y(e),N(r),SB1: \\ 
21-1586 & 21394793+5726427 & K7 & (-16) & 0.9 & w(c) &   ---    & 2.8 & -10.7$\pm$1.9 & 9.4 & Y(e),P(r) \\ 
24-78 & 21394936+5730546 & M2 & (-6) & 0.3 & w &    --- & 2.1: & -21.6$\pm$2.9: & 12.0: & Y(e),P(r) \\ 
92-1162 & 21394974+5746468 & M2 & (-7) & 0.5 & w &   ---    & 2.7: & -12.4$\pm$1.6: & 8.0: & Y(e),Y(r) \\ 
53-1762 & 21395029+5719177 & M0 & (-2) & 0.4 & w &  ---    & 1.9: & -73.1$\pm$2.1: & 8.0: & Y(e),N(r),SB1: \\ 
24-1047 & 21395088+5736168 & --- & --- & --- & w &   ---   & 7.5 & -14.9$\pm$1.3 & 14.7 & Y(r)* \\ 
93-361 & 21395237+5756186 & G1 & -75(-78) & 0.4 & c &         & --- & --- & --- & Y(e) \\
--- & 21395813+5728335 & --- & -15 & --- & c &   370      & 11.1 & -19.0$\pm$0.7 & 11.8 & Y(e),P(r)* \\ 
21-1692 & 21400128+5727184 & M1 & (-5) & 0.5 & w &   ---   & 4.4 & -16.5$\pm$1.2 & 8.9 & Y(e),Y(r) \\ 
21-763 & 21400259+5722090 & M0 & (-3) & 0.4 & w &   ---   & 2.5 & -14.7$\pm$5.5 & 28.3 & Y(e),Y(r) \\ 
24-817 & 21400273+5735050 & K6.5 & (-4) & 0.6 & w &  ---    & 4.2 & -16.4$\pm$2.2 & 15 & Y(e),Y(r) \\ 
21-895 & 21400321+5722505 & K5 & (-1) & 0.5 & w &    ---   & --- & --- & --- & Y(e) \\
21-1762 & 21400924+5727393 & K5 & (-3) & 0.5 & w &   ---   & 6.2 & -11.8$\pm$1.3 & 12.3 & Y(e),Y(r) \\ 
93-720 & 21400999+5800036 & --- & -16 & --- & c &   470      & 10.3 & -6.2$\pm$0.8 & 12.5 & Y(e),PN(r)* \\ 
24-382 & 21401023+5732511 & K7.5 & (-5) & 0.4 & w &    ---   & 6.2 & -13.1$\pm$0.9 & 8.4 & Y(e),Y(r) \\ 
24-1736 & 21401134+5739518 & M1 & -33(-61) & --- & c &   380      & 1.9: & -11.4$\pm$1.8: & 6.9: & Y(e),Y(r) \\ 
24-1796 & 21401182+5740121 & K7 & -73(-124) & 0.3 & c &   410      & 1.4: & -12.9$\pm$4.1: & 13.2: & Y(e),Y(r),SB2: \\
--- & 21401438+5740507 & --- & -46 & --- & c &    470     & 2.6 & -12.2$\pm$1.8 & 8.7 & Y(e),Y(r)* \\ 
22-2651 & 21402130+5726579 & M1.5 & -48(-34) & --- & c &    470     & --- & --- & --- & Y(e) \\
--- & 21402192+5730054 & K6 & -4(-8) & 0.5 & c(w) &   510      & 9.3 & -16.0$\pm$0.8 & 10.6 & Y(e),Y(r),TOa \\ 
92-1103 & 21402274+5746240 & K5.5 & (-7) & 0.9 & c: &    ---   & 3.1 & -13.6$\pm$1.6 & 9 & Y(e),Y(r) \\ 
22-1418 & 21402287+5727329 & M1.5 & (-2) & 0.5 & c(w) &   ---    & 4.2 & -13.9$\pm$1.6 & 10.7 & Y(e),Y(r) \\ 
92-582 & 21402592+5743271 & --- & --- & --- & w &   ---    & 4.6 & -5.1$\pm$1.1 & 8.6 & PN(r) \\ 
23-405 & 21403134+5733417 & K5 & -9(-14) & 0.4 & c &    410     & 6.9 & -19.3$\pm$1.2 & 12.8 & Y(e),P(r) \\ 
23-570 & 21403574+5734550 & K6 & -18(-47) & 0.4 & c &   520      & 9.5 & -14.5$\pm$0.6 & 8.2 & Y(e),Y(r) \\ 
93-540 & 21403586+5758130 & M0 & -5 (-18) & 0.3 & c &   290    & 2.9 & -13.2$\pm$1.5 &7.6 & Y(e),Y(r) \\
53-1843 & 21403592+5719398 & M0.5 & (-4) & 0.6 & w &    ---  & 3.1 & -14.9$\pm$2.1 & 11.4 & Y(e),Y(r) \\ 
93-168 & 21404061+5754064 & K6.5 & -12(-22) & 0.5 & c &   350      & 6.9 & -13.5$\pm$0.8 & 8.3 & Y(e),Y(r) \\ 
92-926 & 21404150+5745219 & --- & --- & --- & w &    ---  & 12 & -11.0$\pm$0.4 & 6.9 & P(r)* \\ 
23-162 & 21404450+5731314 & K7 & (-6) & 0.4 & c:(w) &   ---   & 5.6 & -14.9$\pm$0.8 & 7.4 & Y(e),Y(r) \\ 
23-753 & 21405391+5736199 & M0.5-M0 & (-2) & 0.5 & w &   ---   & 2.8 & -12.0$\pm$1.7 & 8.3 & Y(e),Y(r) \\ 
53-1561 & 21405592+5717591 & K6 & -32(-28) & 0.1: & c &   440      & 3.3 & -11.0$\pm$1.5 & 8.1 & Y(e),P(r) \\ 
22-445 & 21405869+5721095 & M0.5 & (-4) & 0.4 & w &     ---  & 2.4 & -9.0$\pm$3.6 & 19.9 & Y(e),P(r) \\ 
22-1526 & 21410212+5728220 & M1 & (-1) & 0.5: & w &  ---     & 3.2 & -15.8$\pm$1.8 & 9.9 & Y(e),Y(r) \\ 
23-1282 & 21410550+5741032 & --- & --- & --- & w &   ---   & 7 & -17.0$\pm$0.8 & 8.8 & Y(r)* \\ 
23-969 & 21411497+5738149 & K5.5 & -26(-36) & 0.5 & c &  460       & 8.8 & -14.0$\pm$0.8 & 10.9 & Y(e),Y(r) \\ 
22-960 & 21411523+5724256 & M2.5 & (-3) & 0.4: & w &   ---    & --- & --- & --- & P(e) \\
22-1569 & 21411837+5728433 & M1 & (-3) & 0.6 & w &  ---    & 6.3 & -14.9$\pm$0.8 & 7.5 & Y(e),Y(r) \\ 
23-798 & 21412864+5736432 & K6 & -70(-150) & 0.5 & c &   470      & 1.8: & -11.6$\pm$1.6: & 6.1: & Y(e),Y(r) \\ 
23-259 & 21413235+5732245 & --- & --- & --- & w &   ---    & 9.2 & -11.0$\pm$0.5 & 7.2 & P(r)* \\ 
\enddata 
\tablecomments{Spectroscopic Information. For comparison, together with the parameters derived from the 
high-resolution Echelle spectra included in this work, we include the spectral types and Li 
EW from low-resolution spectra (Paper I, II). The mark ``:'' denotes uncertain
values. H$\alpha$ EW in parentheses refer to the 
values obtained from low-resolution spectra. We define the H$\alpha$ type as ``c'' 
(standing for CTTS) if broad emission with extended velocity wings was detected, otherwise we name it ``w'' 
(standing for WTTS). Whenever the classification based
on the H$\alpha$ EW from low-resolution spectra using the criteria in White \& Basri 2003 differed from
the classification according to line broadening, we include the EW-based result in parentheses
(see Sicilia-Aguilar et al. Paper A for more detailed information). The error in the rotational
velocity is proportional to the rotational velocity times 1/(1+R), where R indicates the goodness of the
cross-correlation. Membership is stated as follows: ``Y'' = sure member; ``P'' = probable member, ``PN'' = probable
non-member, ``N'' = non-member. The label in parentheses indicates whether the membership was obtained
from the presence of H$\alpha$ strong emission and/or Li absorption lines (``e'') or via the radial velocity (``r''). 
A star symbol was added to mark the new members. The label ``conf'' denotes previous possible members (from
low-resolution spectroscopy) that are confirmed as members based on the radial velocities; ``rej'' denotes
previous possible members that are most likely non-members according to radial velocities.
Finally, a comment on binarity (SB2 or SB1) is added in this field.}
\end{deluxetable}
\end{center}

\clearpage

\pagestyle{plaintop}

\begin{deluxetable}{lcccccl}
\tabletypesize{\scriptsize}
\tablenum{2}
\tablecolumns{7} 
\tablewidth{0pc} 
\tablecaption{Summary of Disk and Stellar Properties of Members and Probable Members\label{summary-table}} 
\tablehead{
 \colhead{ID} & \colhead{Sp.Type} &\colhead{Age (Myr)} & \colhead{Mass (M$_\odot$)} & \colhead{V$sini$ (km/s)}
&\colhead{\.{M} (10$^{-8}$M$_\odot$/yr)} & \colhead{Class and Comments} }
\startdata
74-48 & ---  & 2.4 & 0.9 & 7.6$\pm$1.5 & 0 &  III      \\
73-758 & K6.5 & 1.8 & 0.8 & 7.7$\pm$2.1 & 0 & Transition     \\
72-1427 & M1 & 2.2 & 0.5 & --- & $<$0.1 & II,Ra       \\
81-541 & K5.5 & 2.8 & 1 & --- & $<$0.1 & II,Ra       \\
64-156 & ---  & 15.6 & 1.3 & 7.4$\pm$1.6 & 0 &  III      \\
73-472 & K5 & 3.3 & 1.1 & 9.6$\pm$2.2 & $<$0.1 &  III      \\
73-1059 & ---  & 14.4 & 1 & 8.6$\pm$1.4 & 0 &  III      \\
73-311 & M1.5 & 2.8 & 0.4 & --- & $<$0.1 & II,B:       \\
73-71 & K6 & 1.8 & 0.9 & 24.4$\pm$5.2 & 2.1 & II,Ra,Ba       \\
72-875 & M0.5 & 8.4 & 0.5 & 7.0$\pm$2.7 & 0 & II:,SB1:      \\
61-608 & ---  & 10.6 & 1.2 & 8.2$\pm$1.5 & 0 &  III      \\
73-194 & K6.5 & 14.3 & 0.8 & 7.3$\pm$1.7 & 0 &  III      \\
61-893 & ---  & 9.4 & 0.9 & 9.9$\pm$3.0: & 0 &  III      \\
21360745+5734296 & ---  & --- & --- & 10.2$\pm$4.2 & 0 & III:      \\
73-537 & G1.5 & 27.4 & 1.3 & 11.7$\pm$2.9 & 0 &  III      \\
84-23 & ---  & 4 & 1 & 9.6$\pm$2 & 0 &  III      \\
21362368+5732452 & ---  & --- & --- & 7.7$\pm$3.0: & 0 & III:      \\
21362507+5727502 & M0 & --- & --- & 7.6$\pm$2.2 & $\dots$ & II,SB1,B       \\
14-306 & K6.5 & 12.8 & 0.9 & --- & 0 &  III      \\
21364596+5729339 & --- & --- & --- & 5.7$\pm$2.4: & $\dots$ &  I      \\
21364762+5729540 & K6 & --- & --- & --- & 0 & III:      \\
14-141 & K6 & 0.6 & 0.8 & 27.0$\pm$8.3: & $<$0.1 & II,Ba       \\
14-1229 & K6 & 10.7 & 0.9 & --- & 0 &  III      \\
11-2146 & K6 & 0.9 & 0.9 & 12.6$\pm$1.5 & 16.2-13.2 & II,Za       \\
11-1209 & K6 & 1 & 0.8 & 12.2$\pm$1.1 & 0 & II,B       \\
21365947+5731349 & M0 & --- & --- & --- & $\dots$ & II,Ba       \\
11-1659 & K5 & 5 & 1.1 & 9.1$\pm$1 & 0.1273 & III:      \\
11-1499 & M1.5 & 1.4 & 0.4 & --- & 0 &  III      \\
11-2322 & M1 & 0.8 & 0.5 & 9.5$\pm$2.6: & 0.80 & II,Ba       \\
11-1871 & M2 & 2.5 & 0.4 & --- & 0.12: & III:      \\
14-222 & K7 & 0.7 & 0.7 & --- & 0.61-0.86 & Transition:      \\
14-287 & M0 & 1 & 0.6 & 11.0$\pm$2.8 & $<$0.1 & II       \\
11-2037 & K4.5 & 2.5 & 1.3 & 8.6$\pm$0.8 & 0.97-2.5 & II       \\
11-1067 & M0.5 & 2.9 & 0.5 & --- & 0 &  III      \\
14-11 & M1.5 & 0.7 & 0.5 & 15.1$\pm$5 & 0 & Transition      \\
14-125 & K5 & 2.9 & 1.1 & 5.6$\pm$1.5 & $<$0.1 & II,B       \\
11-1513 & K7.5 & 1.2 & 0.6 & --- & 0 &  III      \\
11-2131 & K6.5 & 2.3 & 0.8 & --- & 1.1 & II,IPC       \\
11-2487 & K7 & 2.8 & 0.8 & --- & 0 &  III      \\
11-2031 & K2 & 2.5 & 1.7 & 12.6$\pm$1.6 & 1.6 & II       \\
14-103 & K7 & 14.4 & 0.8 & --- & 0 &  III      \\
14-197 & K5.5 & 4.2 & 1 & 20.4$\pm$6.5: & 0 &  III,SB1:      \\
21372410+5724115 & ---  & --- & --- & --- & $\dots$ & II,S       \\
14-160 & K5 & 3.1 & 1.1 & 22.5$\pm$4.2 & $<$0.1 & II       \\
14-1017 & M0 & 2.2 & 0.5 & --- & $<$0.1 & II,Ra       \\
14-335 & K6.5 & 2.8 & 0.8 & --- & $\dots$ & II       \\
11-1864 & G-K & 7.4 & 1 & 9.5$\pm$2.3 & 0 & III:,SB1      \\
83-343 & M0.5 & 0.9 & 0.5 & 8.8$\pm$2 & $\dots$ & II :      \\
14-183 & K7.0(K5)R & 0.9 & 0.8 & 12.8$\pm$4 & $<$0.1 & II,SB1:,Ba       \\
14-2148 & M1.5 & 2 & 0.4 & 4.8$\pm$1.7: & 0 & Transition:      \\
21374275+5733250 & F9 & --- & --- & --- & $\dots$ & II,Ba       \\
11-1384 & K6.5 & 2.4 & 0.8 & --- & 0 &  III      \\
11-383 & K5 & 4.4 & 0.9 & --- & $<$0.1 & II,S       \\
21374893+5723209 & ---  & --- & --- &  --- & $\dots$ & II,Ba       \\
13-924 & K5 & 4.7 & 1.1 & 8.8$\pm$2 & 0 &  III      \\
12-1984 & K6 & 5.8 & 0.9 & --- & 0 &  III      \\
12-2519 & K5.5 & 6.2 & 1 & 10.5$\pm$2.9 & $<$0.1 & II       \\
12-1968 & K6 & 2.6 & 0.9 & 10.9$\pm$1.4 & $<$0.1 & II+III:, SB2:     \\
12-1422 & M0 & 2.7 & 0.5 & --- & 0 &  III      \\
12-1091 & G2.5 & 23.4 & 1.4 & 8.8$\pm$1.2 & 0.81-3.3 & II,Ba      \\
13-269 & K6.5 & 1.6 & 0.8 & --- & 0 &  III      \\
12-583 & M0 & 0.9 & 0.5 & --- & 0 &  III      \\
13-1143 & ---  & 9.6 & 1.1 & 7.2$\pm$1.1 & 0 &  III      \\
13-1238 & M1 & 0.8 & 0.5 & --- & 10.7 & II,Ba       \\
12-2373 & M1 & 1.8 & 0.5 & 10.0$\pm$2.8 & 0 &  III      \\
82-272 & G9 & 10.5 & 1.5 & 18.1$\pm$5 & 23.9 & II+II,SB2,S      \\
12-1081 & M0.5 & 4 & 0.5 & --- & 0 &  III      \\
13-1161 & M0 & 2.8 & 0.6 & 12.8$\pm$4.0: & 0 &  III      \\
12-1613 & M1 & 2.3 & 0.4 & --- & 0 &  III      \\
13-1426 & M0 & 1.8 & 0.5 & --- & $<$0.1 & II,Ba,Za       \\
21380924+5720198 & ---  & --- & --- & 8.2$\pm$3.2 & $\dots$ &  II,S     \\
13-669 & K1 & 2.8 & 1.8 & 14.2$\pm$1.4: & 0.59-2.2 & II,S       \\
21380979+5729428 & ---  & 21380979 & 0 & 9.0$\pm$2.3 &  & II       \\
13-350 & M1 & 4.2 & 0.5 & 6.4$\pm$1.7 & 0 & Transition:      \\
12-1017 & K5.5 & 4 & 1 & 8.8$\pm$0.8 & 0.06 &  III      \\
54-1781 & M1 & 2.3 & 0.4 & 7.1$\pm$2.3 & $<$0.1 & II       \\
13-1877 & K7 & 1 & 0.8 & --- & 1.9-2.7 & II,Ba       \\
13-277 & G1 & 2.6 & 2.6 & 51.6$\pm$9.4 & 30: & II,SB1:,Ba  \\
13-2236 & K6.5 & 5.2 & 0.9 & 15.2$\pm$2.4 & 0 &  III      \\
12-1009 & K5.5 & 2.9 & 1 & 12.6$\pm$2.7 & 0 &  III      \\
94-1119 & ---  & 2.6 & 0.9 & 7.3$\pm$0.6 & 0 &  III      \\
13-819 & K5.5 & 2.4 & 1 & --- & 0.04-0.20 & Transition     \\
94-1050 & ---  & 16.2 & 0.9 & 7.8$\pm$2.6 & 0 &  III      \\
12-1955 & K6.5 & 6.9 & 0.9 & 9.0$\pm$1.1 & 0 &  III      \\
13-236 & K2 & 2.8 & 1.7 & 9.0$\pm$0.9 & 1.5-3.9 & II,S       \\
12-2113 & K6 & 1.1 & 0.9 & 11.7$\pm$2.5 & 6 & II,SB1,S   \\
13-157 & K5.5 & 2.4 & 1 & 8.7$\pm$1.8 & 1.6 & II,Ba       \\
13-232 & M0 & 2.2 & 0.6 & --- & 0 & III:      \\
21383216+5726359 & M0 & --- & --- & --- & $\dots$ & II       \\
13-52 & K7 & 4.1 & 0.8 & 8.1$\pm$1.9 & 0 & Transition     \\
12-1825 & ---  & 4.8 & 0.5 & 8.8$\pm$2.6 & 0 &  III      \\
91-155 & M2.5 & 1.7 & 0.4 & 5.9$\pm$2.4 & $\dots$ & II       \\
13-566 & K5.5 & 2.5 & 0.8 & --- & 0 & Transition:      \\
13-1891 & M0 & 5.1 & 0.5 & 10.1$\pm$2.3 & $\dots$ & II,Za       \\
13-1709 & K5.5 & 5.2 & 1 & 10.4$\pm$2.4 & 0.03 &  III      \\
54-1613 & K5 & 5.2 & 1.1 & 9.5$\pm$0.7 & 0.37 & Transition:      \\
21384350+5727270 & M2 & --- & --- & --- & $\dots$ & Transition      \\
54-1547 & K5.5 & 5.7 & 1 & 20.0$\pm$7.1 & 0.21 & II,Ra       \\
12-2363 & M0.5 & 0.9 & 0.5 & 27.1$\pm$7.3 & 0 &  III,SB1: \\
12-595 & K7 & 18.9 & 0.7 & --- & 0 & Transition:     \\
12-1423 & K7 & 2.3 & 0.8 & 7.5$\pm$1.5 & $<$0.1 &  III      \\
12-1010 & M2 & 5.3 & 0.4 & 8.2$\pm$2.3 & $\dots$ & II,S       \\
12-2098 & M2.5 & 2.2 & 0.4 & --- & 0 &  III      \\
21-851 & ---  & 25.6 & 0.9 & 6.0$\pm$2.4 & 0 &  III      \\
13-1087 & K4 & 4.2 & 1.4 & 11.2$\pm$0.8 & 0.28-0.37: & III:      \\
91-506 & K6.5 & 2.5 & 0.8 & 8.4$\pm$2.1 & 0.14 & II,Ra       \\
24-692 & M1 & 2.7 & 0.5 & 6.0$\pm$1.9 & 0 &  III,SB1  \\
12-1027 & M0 & 2.6 & 0.4 & ---& 0 & III     \\
24-77 & K6.5 & 7.4 & 0.9 & --- & 0 &  III      \\
24-108 & K5.5 & 4.3 & 1 & --- & 0 &  III      \\
12-1617 & M1 & 1.2 & 0.5 & 6.1$\pm$2.3: & 0.88 & II       \\
13-1048 & M0 & 7.4 & 0.6 & 10.9$\pm$1.4 & $<$0.1 & II,Za       \\
13-1250 & K4.5 & 3.3 & 1.3 & 8.3$\pm$0.8 & 0.10 & Transition,IPC     \\
21-563 & M1 & 1.4 & 0.6 & --- & $\dots$ &  III      \\
12-942 & K7.5 & 1.5 & 0.7 & 19.3$\pm$3.8 & 0 &  III      \\
91-815 & M2 & 2 & 0.4 & 6.5$\pm$2.2 & 0 &  III      \\
21-1590 & K7 & 3.4 & 0.8 & 10.6$\pm$2 & 0 &  III      \\
21392541+5733202 & ---  & --- & --- & 8.8$\pm$1 & $\dots$ & Transition,S     \\
21392570+5729455 & ---  & --- & --- & 7.5$\pm$2.3 & $\dots$ & Transition      \\
24-542 & K4 & 2.3 & 1.1 & --- & 0 &  III      \\
21393104+5747140 & ---  & --- & --- & 8.8$\pm$0.5 & $\dots$ & II,S       \\
24-515 & M0.5 & 2.8 & 0.5 & 7.6$\pm$2.6: & $<$0.1 & Transition      \\
21-998 & K5.5 & 5.6 & 1 & 7.3$\pm$2.8 & 0.73 & II,SB1:  \\
21-33 & M0 & 4 & 0.5 & 8.9$\pm$3.1: & $<$0.1 & II,S       \\
24-170 & K7.5 & 2.4 & 0.7 & 7.0$\pm$2.9: & 0 & III:      \\
53-1803 & K6.5 & 20.5 & 0.8 & 7.2$\pm$1.6 & 0 &  III,SB1 \\
24-48 & M0.5 & 2.4 & 0.5 & 8.1$\pm$1.1 & 0 &  III      \\
92-1198 & ---  & 2.3 & 1 & 14.1$\pm$1.6 & 0 &  III      \\
21-230 & M0.5 & 2.5 & 0.5 & 11.3$\pm$2.1 & 0 &  III      \\
92-393 & M2 & 0.9 & 0.4 & 14.2$\pm$4.1 & $<$0.1 & Transition,Za      \\
24-820 & K6.5 & 26.7 & 0.8 & 6.9$\pm$1.4 & 0 &  III      \\
21-2251 & M2 & 1.3 & 0.4 & 16.7$\pm$5.9: & 0 &  III,SB1: \\
21-1586 & K7 & 3.9 & 0.6 & 9.4$\pm$2.5 & 0 &  III      \\
24-78 & M2 & 1.4 & 0.4 & 12.0$\pm$3.8: & 0 &  III      \\
92-1162 & M2 & 1.6 & 0.4 & 8.0$\pm$2.1: & 0 &  III      \\
53-1762 & M0 & 3 & 0.6 & 8.0$\pm$2.8: & 0.434 & II:,SB1: \\
24-1047 & ---  & 2.3 & 1 & 14.7$\pm$1.7: & 0 &  III      \\
93-361 & G1 & 15.6 & 1.5 & --- & $<$0.1 & II,Za       \\
21395813+5728335 & ---  & --- & --- & 11.8$\pm$1 & $\dots$ & II,S      \\
21-1692 & M1 & 2.5 & 0.5 & 8.9$\pm$1.6 & 0 &  III      \\
21-763 & M0 & 2.6 & 0.6 & 28.3$\pm$8.1 & 0 &  III      \\
24-817 & K6.5 & 4.6 & 0.9 & 15.0$\pm$2.9 & 0 &  III      \\
21-895 & K5 & 2.8 & 1.1 & --- & 0: & III:      \\
21-1762 & K5 & 4.3 & 1.1 & 12.3$\pm$1.7 & 0.23 &  III      \\
93-720 & ---  & 1.8 & 1 & 12.5$\pm$1.1 & $\dots$ & II,Ra       \\
24-382 & K7.5 & 6 & 0.7 & 8.4$\pm$1.2 & 0 &  III      \\
24-1736 & M1 & 3.9 & 0.4 & 6.9$\pm$2.4: & 0.18 & II,Ra       \\
24-1796 & K7 & 6 & 0.8 & 13.2$\pm$5.6: & 0.31-0.55 & II+II, SB2:      \\
21401438+5740507 & ---  & --- & --- & 8.7$\pm$2.4 & $\dots$ & II,Ba      \\
22-2651 & M1.5 & 3.9 & 0.4 & --- & $<$0.1 & II,Za      \\
21402192+5730054 & K6 & --- & --- & 10.6$\pm$1 & $\dots$ & Transition,IPC     \\
92-1103 & K5.5 & 1.2 & 1 & 9.0$\pm$2.2 & $<$0.1 & II:      \\
22-1418 & M1.5 & 2.1 & 0.4 & 10.7$\pm$2.1 & $<$0.1 & II:      \\
23-405 & K5 & 5 & 1.1 & 12.8$\pm$1.6 & $<$0.1 & II,Za       \\
23-570 & K6 & 4.2 & 0.9 & 8.2$\pm$0.8 & 0.62 & II,Ba      \\
93-540 & M0 & 2.3 & 0.6 & 7.6$\pm$1.9 & $<$0.1 & II       \\
53-1843 & M0.5 & 3.6 & 0.5 & 11.4$\pm$2.8 & 0 &  III      \\
93-168 & K6.5 & 2.5 & 0.8 & 8.3$\pm$1.1 & 0.05 & II,Za       \\
92-926 & ---  & 6.9 & 1.3 & 6.9$\pm$0.5 & 0 &  III      \\
23-162 & K7 & 6.6 & 0.8 & 7.4$\pm$1.1 & $<$0.1 & II       \\
23-753 & M0.5-M0 & 8.5 & 0.5 & 8.3$\pm$2.2 & 0 &  III      \\
53-1561 & K6 & 3 & 0.8 & 8.1$\pm$1.9 & $\dots$ & II,Za       \\
22-445 & M0.5 & 2.4 & 0.5 & 19.9$\pm$5.9 & 0 &  III      \\
22-1526 & M1 & 3.2 & 0.5 & 9.9$\pm$2.3 & 0 &  III      \\
23-1282 & ---  & 6.7 & 1.1 & 8.8$\pm$1.1 & 0 &  III      \\
23-969 & K5.5 & 2.9 & 1 & 10.9$\pm$1.1 & 0.41 & II,Za       \\
22-960 & M2.5 & 2.5 & 0.4 & --- & 0 &  III      \\
22-1569 & M1 & 1.3 & 0.5 & 7.5$\pm$1 & $\dots$ & III:      \\
23-798 & K6 & 14.7 & 0.9 & 6.1$\pm$2.2: & 0.79 & II,Ba       \\
23-259 & ---  & 2.2 & 1.1 & 7.2$\pm$0.7 & 0 &  III      \\
\enddata
\tablecomments{Summary of the properties of the stars in Tr 37. 
Only the star being sure or probable members are displayed here.
Ages and masses are derived from the V vs. V-I diagram,
using the isochrones by Siess et al. 2000 (see Paper I, Paper II for a detailed
description of spectral typing and age/mass calculation).
Errors in the V$sini$ are derived as V$sini$/(1+R) (see Tonry \& Davis 1979; Hartmann et al.
1986). The best estimate of the
accretion rate is selected for this summary. Whenever accretion rates derived from
U band photometry were available (Paper II), we included them. Since some stars had been
observed in U band in two campaigns, we include both values. As in the text, we denote
with \.{M}=0 those stars with no U excess and H$\alpha$ profiles
consistent with diskless WTTS or Class III objects.
For stars with no detectable U excess but broad profiles, we give an upper limit to
accretion of $<$0.1 10$^{-8}$M$_\odot$/yr (see text).
We do not provide an accretion value for stars not observed in U band
and broad or probably broad H$\alpha$ profiles. For the exceptional star 13-277, whose spectral
type is uncertain due to veiling, we provide an approximate
accretion rate from the H$\alpha$ width at 10\% maximum (see text). Most of these stars without a
value for accretion are likely to be accreting (showing broad profiles, see figures).
The comments include the class (I, II, III or Transition object), the binarity 
(SB1= single-lined spectroscopic
binary; SB2= double-lined spectroscopic binary), and the details on the H$\alpha$ profile (Ba= blue-shifted
absorption; Ra= red-shifted absorption; Za= absorption with approximately zero velocity shift;
B= blue-shifted emission; R=red-shifted emission; IPC= inverse P-Cygni profile).
Uncertain values are denoted by ``:''. }
\end{deluxetable}


\begin{thebibliography}{}

\bibitem[Alexander et al.(2006a)]{alexander06a} Alexander, R.C., Clarke, C.J., Pringle, J.E., 2006a, MNRAS in press

\bibitem[Alexander et al.(2006b)]{alexander06b} Alexander, R.C., Clarke, C.J., Pringle, J.E., 2006b, MNRAS in press

\bibitem[Andr\'{e} \& Montmerle(1994)]{andre94} Andr\'{e}, Ph., \& Montmerle, T., 1994, \apj, 420, 837

\bibitem[Appenzeller \& Mundt(1989)]{appenzeller89} Appenzeller, I., \& Mundt, R., 1898, AARV, 1, 291

\bibitem[Armitage et al.(2003)]{armitage03} Armitage, Ph., Clarke, C., Palla, F., 2003, MNRAS, 342, 1139

\bibitem[Bergin et al.(2004)]{bergin04}Bergin, E., Calvet, N., Sitko, M., and 9 more coauthors, 2004, \apj, 614, L133

\bibitem[Bertout(1989)]{bertout89} Bertout, C., 1989, ARA\&A, 27, 351

\bibitem[Bonnell et al.(1998)]{bonnell98} Bonnell, I.~A., Smith, K.~W., Meyer, M.~R., Tout, C.~A., Folha, D.~F.~M., \& Emerson, J.~P.\ 1998, \mnras, 299, 1013 

\bibitem[Bouvier et al.(1993)]{bouvier93} Bouvier, J., Cabrit, S., Fern\'{a}ndez, M., Mart\'{\i}n, E.L., Matthews, J.M., AA, 272. 176

\bibitem[Bryden et al.(2000)]{bryden00}Bryden, G, Rozyczka, M., Lin, D., Bodenheimer, P., 2000, 540, 1091

\bibitem[Cabrit et al.(1990)]{cabrit90} Cabrit, S., Edwards, S., Strom, S., Strom, K., 1990, \apj, 354, 687

\bibitem[Calvet \& Hartmann(1992)]{calvet92} Calvet, N. \& Hartmann, L. W. 1992, \apj, 386, 239

\bibitem[Calvet \& Gullbring(1998)]{calvet98} Calvet, N. \& Gullbring, E., 1998, \apj, 509,802

\bibitem[Calvet et al.(2002)]{calvet02} Calvet, N., D'Alessio, P., Hartmann, L., Wilner, D., Walsh, A. \& Sitko, M., 2002, \apj, 568, 1008

\bibitem[Calvet et al.(2004)]{calvet04} Calvet, N., Muzerolle, J., Brice{\~n}o, C., Hern{\'a}ndez, J., Hartmann, L., Saucedo, J.~L., \& Gordon, K.~D.\ 2004, \aj, 128, 1294

\bibitem[Calvet et al.(2005)]{calvet05} Calvet, N., Brice\~{n}o, C., Hern\'{a}ndez, J., Hoyer, S., Hartmann, L., Sicilia-Aguilar, A., Megeath, S.T., D'Alessio, P., 2005, \aj 139, 935

\bibitem[Carpenter et al.(1990)]{carpenter90} Carpenter, J., Snell, R., Schloerb, F.P., 1990, \apj, 362, 147

\bibitem[Choi and Herbst(1996)]{choi96} Choi, P.I., \& Herbst, W., 1996, \aj, 111, 283

\bibitem[Clarke et al.(2001)]{clarke01} Clarke, C., Gendrin, A., \& Sotomayor, M., 2001, MNRAS 328, 485

\bibitem[Contreras et al.(2002)]{contreras02}  Contreras, M.E., Sicilia-Aguilar, A., Muzerolle, J., Calvet, N., Berlind, P., Hartmann, L. 2002, AJ, 124, 1585

\bibitem[D'Alessio et al.(2005)]{dalessio05} D'Alessio, P., Hartmann, L., Calvet, N., Franco-Hern\'{an}dez, R., and 10 more coauthors, 2005, \apj, 621, 461

\bibitem[D'Angelo et al.(2003)]{dangelo03} D'Angelo, G., Henning, Th., \& Kley, W., 2003, \apj, 599, 548

\bibitem[Edwards et al.(1993)]{edwards93} Edwards, S., Strom, S., Hartigan, P., Strom, K., Hillenbrand, L., Herbst, W., Attridge, J., Merrill, K., Probst, R., Gatley, I., 1993, AJ, 106, 372

\bibitem[Fabricant et al.(2004)]{fabricant04}Fabricant, D.G.  et al 2004, Proc SPIE in press

\bibitem[Forrest et al.(2004)]{forrest04} Forrest, W.J., Sargent, B., Furlan, E., \& 18 more coauthors, 2004, \apjs, 154, 443

\bibitem[F\"{u}r\'{e}sz et al.(2006)]{furesz06} F\"{u}r\'{e}sz, G., et al., 2006, ApJ in press

\bibitem[Gregorio-Hetem \& Hetem(2002)]{gregoriohetem02} Gregorio-Hetem, J., \& Hetem, A. Jr., 2002, MNRAS, 336, 197

\bibitem[Gullbring et al.(1998)]{gullbring98} Gullbring, E., Hartmann, L., Brice\~{n}o, C., Calvet, N., 1998, ApJ 492, 323

\bibitem[Haisch et al.(2001)]{haisch01} Haisch, K., Lada, E., \& Lada, C., 2001, \apj, 553,153

\bibitem[Hartigan et al.(1990)]{hartigan90} Hartigan, P., Hartmann, L., Kenyon, S., Strom, S., \& Skrutskie, M., 1990, \apj, 354, 25

\bibitem[Hartmann et al.(1986)]{hartmann86} Hartmann, L., Hewett, R., Stahler, S., Mathieu, R., 1986, \apj, 309, 275

\bibitem[Hartmann \& Kenyon(1996)]{hartmann96} Hartmann, L., \& Kenyon, S., 1996, ARAA, 34, 207

\bibitem[Hartmann(1998)]{hartmann98} Hartmann, L.: Accretion Processes in Star Formation, Cambridge University Press, 1998.

\bibitem[Hartmann et al.(1998)]{hartal98} Hartmann, L., Calvet, N., Gullbring, E. \& D'Alessio, P, 1998, \apj, 495, 385

\bibitem[Hartmann(2002)]{hartmann02} Hartmann, L., 2002, \apj, 566, 29

\bibitem[Hartmann(2003)]{hartmann03} Hartmann, L., 2003, \apj, 585, 398

\bibitem[Herbig(1998)]{herbig98} Herbig, G.H., 1998, \apj, 497, 736

\bibitem[Herbst et al.(1994)]{herbst94} Herbst, W., Bailer-Jones, C., Mundt, R., Meisenheimer, K., Wackermann, R., 2002, AA, 396, 513 

\bibitem[Herbst et al.(2002)]{herbst02} Herbst, W., Herbst, D.K., Grossman, E.J, \& Weinstein, D., 1994, \aj, 108, 1906

\bibitem[Kenyon \& Hartmann(1995)]{kenyon95}  Kenyon, S.J. \& Hartmann, L., 1995, ApJS , 101, 117

\bibitem[K\"{u}ker et al.(2003)]{kuker03} K\"{u}ker, M., Henning, Th., R\"{u}diger, G., 2003, \apj, 589, 397

\bibitem[Kurtz et al.(1992)]{kurtz92} Kurtz, M.J., Mink, D.J., Wyatt, W.F., Fabricant, D.G., Torres, G., 
Kriss, G., and Tonry, J.L. 1992, in Astronomical Data Analysis Software and Systems I, 
ASP Conf. Ser., Vol. 25, eds. D.M. Worral, C. Biemesderfer, and J. Barnes, 432

\bibitem[Lada(1987)]{lada87} Lada, C.J., 1987, IAUS, 115, 1L

\bibitem[Lada et al.(2000)]{lada00} Lada, C.J., Muench, A.A., Haisch, K.E., Lada, E.A., Alves, J.F., Tollestrup, E.V. \& Willner, S.P., 2000, \aj, 120, 3162

\bibitem[Lin and Papaloizou(1986)]{lin86} Lin, D.N.C., \& Papaloizou, J., 1986, \apj, 309, 846 
 
\bibitem[Littlefair et al.(2004)]{littlefair04} Littlefair, S., Naylor, T., Harries, T., Retter, A., O'Toole, S., 2004, MNRAS 347, 937

\bibitem[Mathieu et al.(1995)]{mathieu95} Mathieu, R., Adams, F., Fuller, G., et al. 1995, AJ 109, 265

\bibitem[Morgenroth(1939)]{morgenroth39} Morgenroth, O., 1939, Astron. Nach. 268, 273

\bibitem[Muzerolle et al.(1998a)]{muzerolle98a} Muzerolle, J., \& Hartmann, L., \& Calvet, N., 1998a, AJ, 116, 455

\bibitem[Muzerolle et al.(1998b)]{muzerolle98b} Muzerolle, J., Calvet, N. \& Hartmann, L., 1998b, \apj, 492, 743

\bibitem[Muzerolle et al.(2000)]{muzerolle00} Muzerolle, J., Calvet, N., Brice\~{n}o, C., Hartmann, L. \& Hillenbrand, L., 2000, \apj, 535, L47

\bibitem[Muzerolle et al.(2001)]{muzerolle01} Muzerolle, J., Calvet, N. \& Hartmann, L., 2001, \apj, 550, 944

\bibitem[Muzerolle et al.(2003)]{muzerolle03} Muzerolle, J., Hillenbrand, L., Calvet, N., Brice\~{n}o, C., Hartmann, L., 2003, \apj, 592,266

\bibitem[Natta et al.(2004)]{natta04} Natta, A., Testi, L., Muzerolle, J., Randich, S., Comeron, F., \& Persi, P., 2004, A\&A, 424, 603

\bibitem[Patel et al.(1995)]{patel95}  Patel, N.A. , Goldsmith, P.F., Snell, R.L., Hezel, T. \& Xie, T., 1995, ApJ , 447, 721

\bibitem[Patel et al.(1998)]{patel98} Patel, N.A.,  Goldsmith, P.F., Heyer, M.H. \& Snell, R.L., 1998, ApJ , 507, 241

\bibitem[Platais et al.(1998)]{platais98} Platais, I,  Kozhurina-Platais, V., van Leeuwen, F., 1998, \aj, 116, 2423

\bibitem[Podosek \& Cassen(1994)]{podosek94} Podosek, F.A. \&  Cassen, P., 1994, Meteoritics 29, 6-25

\bibitem[Quillen et al.(2004)]{quillen04} Quillen, A., Blackman, E., Frank, A., Varniere, P., 2004, \apj, 612, L137

\bibitem[Randich et al.(2001)]{randich01} Randich, S., Pallavicini, R., Meola,G., Stauffer, J.R., \&  Balachandran, S.C., 2001, \aap, 372, 862

\bibitem[Reipurth et al.(1996)]{reipurth96} Reipurth, B., Pedrosa, A., Lago, M.T.V.T., 1996., AASS, 120, 229

\bibitem[Rhode et al.(2001)]{rhode01} Rhode, K.,  Herbst, W. \&  Mathieu, R., 2001, \aj, 122, 3258

\bibitem[Shu et al.(1994)]{shu94} Shu, F., Najita, J., Ostriker, E., Wilkin, F., Ruden, S., \& Lizano, S., 1994, \apj, 429, 781

\bibitem[Sicilia-Aguilar et al.(2004)]{sic04} Sicilia-Aguilar, A., Hartmann, L., Brice\~{n}o, C., Muzerolle, J., \& Calvet, N., 2004, \aj 128, 805, Paper I

\bibitem[Sicilia-Aguilar et al.(2005)]{sic05onc} Sicilia-Aguilar, A., Hartmann, L., Szentgyorgyi, A., Roll, J., 
Conroy, M., Calvet, N., Fabricant, D., \& Hern\'{a}ndez, J.,  2005, AJ, 129, 363 

\bibitem[Sicilia-Aguilar et al.(2005))]{sic05opt} Sicilia-Aguilar, A., Hartmann, L., Hern\'{a}ndez, J., Brice\~{n}o, C., Calvet, N., 2005, \aj 130, 188, Paper II

\bibitem[Sicilia-Aguilar et al.(2006))]{sic05ir} Sicilia-Aguilar, A., Hartmann, L., Calvet, N., Megeath, S.T., Muzerolle, J., Allen, L., D'Alessio, P., Mer\'{\i}n, B., Stauffer, J., Young, E., Lada, C., 2006, ApJ 638, 897, Paper III

\bibitem[Siess et al.(2000)]{siess00} Siess, L., Dufour, E. \&  Forestini, M. 2000 A\&A , 358,593S

\bibitem[Skrutskie et al.(1990)]{skrutskie90} Skrutskie, M., Dutkevitch, D., Strom, S., Edwards, S., \& Strom, K., 1990, \aj, 99, 1187

\bibitem[Spitzer(1984)]{spitzer84} Spitzer, L., 1984, Sci. 225, 466

\bibitem[Stassun et al.(2001)]{stassun01} Stassun, K., Mathieu, R., Vrba, T., Henden, A., 2001, AJ, 121, 1003

\bibitem[Stassun et al.(1999)]{stassun99} Stassun, K., Mathieu, R., Mazeh, T.,  Vrba, T., 1999, AJ, 117, 2941

\bibitem[Strom et al.(1989)]{strom89} Strom, K., Wilkin, F., Strom, S., \& Seaman, R., 1989, \aj, 98, 1444

\bibitem[Szentgyorgyi et al.(1998)]{sze98} Szentgyorgyi, A.~H., Cheimets, P., Eng, R., Fabricant, D.~G., Geary, J.~C., Hartmann, L., 
Pieri, M.~R., \& Roll, J.~B.,  1998, \procspie, 3355, 242 

\bibitem[Tonry \& Davis(1979)]{tonry79} Tonry, J., \& Davis, M., 1979, AJ 84, 1511

\bibitem[Uchida et al.(2004)]{uch04} Uchida, K.~I., et al. 2004, \apjs, 154, 439

\bibitem[Vittone \& Errico(2005)]{vittone05} Vittone, A., \& Errico, L., 2005, Mem. S.A.It. 76, 320

\bibitem[White \& Basri(2003)]{white03} White, R., \& Basri, G., 2003, \apj, 582, 1109

\end{thebibliography}
\end{document}